\documentclass[aps,preprint,superscriptaddress]{revtex4-1}

\usepackage{xcolor}
\usepackage{amsmath}
\usepackage{graphicx}
\usepackage{bm} 
\usepackage{slashed}
\usepackage{units}
\usepackage{subfigure}
\usepackage{amsfonts}
\usepackage{hyperref}
\hypersetup{
    colorlinks,
    linkcolor={red!50!black},
    citecolor={blue!50!black},
    urlcolor={blue!80!black}
}

\usepackage{ulem}

\date{\today}
\begin{document}

\title{Electric dipole moments of three-nucleon systems in the pionless effective field theory}

\author{Zichao Yang}
\email{zyang32@vols.utk.edu}
\affiliation{Department of Physics and Astronomy, University of Tennessee, Knoxville, TN 37996, USA}
\author{Emanuele Mereghetti}
\email{emereghetti@lanl.gov}
\affiliation{Theoretical Division, Los Alamos National Laboratory, Los Alamos, NM 87545, USA}
\author{Lucas~Platter}
\email{lplatter@utk.edu}
\affiliation{Department of Physics and Astronomy, University of Tennessee, Knoxville, TN 37996, USA}
\affiliation{Physics Division, Oak Ridge National Laboratory, Oak Ridge, TN 37831, USA}
\author{Matthias R.~Schindler}
\email{mschindl@mailbox.sc.edu}
\affiliation{Department of Physics and Astronomy, University of South Carolina, Columbia, SC 29208, USA}
\author{Jared Vanasse}
\email{jvanass3@fitchburgstate.edu}
\affiliation{Fitchburg State University, Fitchburg MA 01420, USA}
\date{\today}

\begin{abstract}
  We calculate the electric dipole moments (EDMs) of three-nucleon
  systems at leading order in pionless effective field theory.  The
  one-body contributions that arise from permanent proton and neutron
  EDMs and the two-body contributions that arise from CP-odd
  nucleon-nucleon interactions are taken into account.  Neglecting the
  Coulomb interaction, we consider the triton and ${}^3$He, and also
  investigate them in the Wigner-SU(4) symmetric limit. We also
  calculate the electric dipole form factor and find numerically that
  the momentum dependence of the electric dipole form factor in the
  Wigner limit is, up to an overall constant (and numerical accuracy),
  the same as the momentum dependence of the charge form factor.
\end{abstract}

\preprint{LA-UR-20-28735}

\maketitle

\section{Introduction}
\label{sec:into}

The breaking of the discrete symmetries of charge conjugation $C$ and
charge conjugation and parity $CP$ is a necessary condition for the
dynamical generation of a matter-antimatter asymmetry in the Universe
\cite{Sakharov:1967dj}.  In the Standard Model (SM) of particle
physics, $C$ is maximally broken by the different gauge interactions
of left- and right-handed quarks and leptons. The breaking of $CP$ is
much more subtle. In the SM with three generations of quarks, $CP$ is
broken by the phase of the Cabibbo-Kobayashi-Maskawa (CKM) mixing
matrix \cite{Kobayashi:1973fv} and by the QCD $\bar\theta$ term
\cite{tHooft:1976rip,tHooft:1976snw}. While all observed $CP$
violation (CPV) in the kaon and $B$ meson systems can be explained by
the CKM mechanism, CPV in the SM fails to generate the observed
matter-antimatter asymmetry of the Universe by several orders of
magnitude
\cite{Gavela:1993ts,Gavela:1994ds,Gavela:1994dt,Huet:1994jb}. Baryogenesis
thus requires the existence of new sources of CPV.

Electric dipole moments (EDMs) of leptons, nucleons, atomic and
molecular systems receive negligible contributions from the CKM
mechanism
\cite{Pospelov:2005pr,Seng:2014lea,Yamanaka:2016fjj,Yamanaka:2015ncb}
and are thus extremely sensitive probes of CPV beyond the SM (BSM).
Currently, the best limits are on the electron EDM,
$|d_e| < 1.1 \cdot 10^{-16}$ $e$ fm (90\% C.L.), deduced from
experiments with ThO and HfF molecules
\cite{Andreev:2018ayy,Cairncross:2017fip}, on the neutron EDM,
$|d_n| < 1.8 \cdot 10^{-13}$ $e$ fm (90\% C.L.) \cite{Abel:2020gbr},
and on the EDM of $^{199}$Hg,
$|d_{^{199}\rm Hg}| < 6.2 \cdot 10^{-17}$ $e$ fm
\cite{Graner:2016ses}.  Constraints on the diamagnetic atoms
$^{129}$Xe and $^{225}$Ra are presently weaker
\cite{Bishof:2016uqx,Sachdeva:2019rkt}, but, particularly in the case
of $^{225}$Ra, they are expected to improve by several orders of
magnitude in the coming years \cite{Bishof:2016uqx}.  These bounds can be
naively converted into new physics scales in the range of $10 - 100$
TeV, making EDM experiments extremely competitive with direct searches
at the Large Hadron Collider (LHC). For this reason, there exists an
extensive experimental program with the goal of improving existing
bounds by one or two orders of magnitude and to search for EDMs in new
systems.  In particular, there are proposals to measure the EDMs of
charged particles, including muons, protons and light nuclei, in
dedicated storage ring experiments
\cite{Orlov:2006su,Pretz:2013us,Abusaif:2019gry,Talman:2018dgp}. 
These experiments might reach a sensitivity of $10^{-16}$ $e$ fm, comparable with the
next generation of neutron EDM experiments, and they provide a much
more direct connection with the microscopic sources of CPV compared to
EDMs of diamagnetic atoms, whose interpretation is affected by the
large nuclear theory uncertainties in the calculations of nuclear Schiff
moments \cite{Ban:2010ea,Engel:2013lsa}.  Thus, the measurement of the
EDMs of the proton and light nuclei might play a crucial role not only
for the discovery of BSM physics, but also in disentangling different
high-energy mechanisms of CPV
\cite{deVries:2011an,Dekens:2014jka,Bsaisou:2014zwa}.

A description of EDM observables that employs nuclear degrees of
freedom is therefore clearly needed for the interpretation of
experimental data. Chiral effective field theory is
particularly useful in this endeavor since it can relate measured EDMs
to their underlying sources, such as the QCD $\bar\theta$-term or CPV
operators from BSM physics. In Weinberg's power counting, the EDMs are
for several BSM mechanisms dominated by pion-range CPV interactions
\cite{deVries:2011an,Bsaisou:2014zwa}, whose strength is related by
chiral symmetry to nucleon masses and mass splittings
\cite{Crewther:1979pi,Mereghetti:2010tp,deVries:2011an,Bsaisou:2014zwa,Seng:2016pfd,deVries:2016jox}.
The CPV pion-nucleon couplings appearing at leading order (LO) can
thus be extracted from existing lattice QCD calculations in the case
of the QCD $\bar\theta$-term \cite{deVries:2015una}, or require
relatively simple lattice QCD input in the case of BSM operators
\cite{deVries:2016jox}. Over the last years significant efforts have
been made to improve the description of EDMs in chiral EFT, with the
derivation of the chiral Lagrangian at next-to-next-to-leading order
(N$^2$LO) from the QCD $\bar\theta$-term and dimension-six sources of
CPV \cite{Mereghetti:2010tp,deVries:2012ab,Bsaisou:2014oka}, and of
the N$^2$LO time-reversal ($T$) breaking potential
\cite{Maekawa:2011vs,Bsaisou:2012rg,Gnech:2019dod,deVries:2020iea}.
These developments made it possible to carry out chiral effective
theory calculations of EDMs of light nuclei
\cite{Bsaisou:2014zwa,Gnech:2019dod}. Such calculations, which employ
a complete effective field theory approach to calculate the wave
function of the nuclear bound state and for the construction of the
nuclear current, promise to provide reliable uncertainty estimates and
a path to the reduction of those quantified uncertainties.  We however stress
that, even in chiral EFT, a systematic connection between nuclear EDMs
and their microscopic quark-level sources beyond LO requires the
determination of CPV nucleon-nucleon couplings, and thus lattice QCD
simulations in two- or three-nucleon systems.  In addition, it was
recently shown in Ref. \cite{deVries:2020loy} that long-standing
issues with the renormalization of singular chiral EFT potentials
\cite{Kaplan:1996xu,Nogga:2005hy} demand the inclusion of LO CPV
short-range nucleon-nucleon couplings whenever the CPV pion-nucleon
interactions act in the $^1S_0$--$^3P_0$ channel. While this has no
consequence for the EDM of the deuteron, it significantly affects the chiral
EFT uncertainties in the three-nucleon system \cite{deVries:2020loy}.

The so-called pionless EFT,
EFT($\slashed{\pi}$)~\cite{Hammer:2019poc}, is an alternative EFT
approach to light nuclei. It is an expansion in the ratio of the range
of the nuclear interaction $R$ and the two-nucleon scattering length
$a$ and has been shown to be a working, order-by-order renormalizable
framework for two-, three- and four-nucleon system
\cite{Kaplan:1998tg, Bedaque:1998km, Bedaque:1998kg,
  Platter:2004he}. The low-energy constants of this EFT can be related
directly to scattering and bound state observables in few-nucleon
systems and pionless EFT predictions are thereby inherently tied to a
small number of these nuclear observables. The dependence of
observables on the chosen regulators is also well-understood and
indicates that the inherent uncertainties of the low-energy expansion
are under control. This EFT can be applied to any system that displays
a large scattering length $a$ and has therefore also found
applications in atomic and particle physics.

Here we will use pionless EFT to calculate the EDM and
the electric dipole form factor (EDFF) of the three-nucleon systems at
leading order. This has several benefits: We can easily study the
dependence of the EDFF on two- and three-nucleon
observables. Furthermore, a non-zero EDM measurement can be directly
related to a corresponding scattering amplitude using pionless EFT.
We can thus retain predictive power by matching these amplitudes to
chiral EFT, at least in those channels in which the CPV pion-exchange
leads to regulator-independent results, or, even more promisingly, by
taking advantage of the significant progress in lattice QCD
calculations of few-nucleon matrix elements
\cite{Nicholson:2015pys,Chang:2017eiq,Horz:2020zvv,Davoudi:2020ngi},
which can be directly related to the corresponding pionless EFT ones.

The paper is organized as follows. In
Sec.~\ref{sec:theor-build-blocks}, we summarize the theoretical
building blocks and define the CPV interactions used in the
calculation. The calculation of the EDFF is conveniently performed by
introducing a trimer field, following
Ref.~\cite{Hagen:2013xga,Vanasse:2015fph,Vanasse:2017kgh}. We give the
integral equation for the CP-even trimer-nucleon-dimer vertex function
in Sec.~\ref{sec:Gfunc}, and derive the integral equations in the
presence of CPV interactions in Sec.~\ref{ToddV}. In
Sec.~\ref{sec:three-nucleon-form}, we give the schematic diagrammatic
expressions of the three-nucleon EDFF, leaving the detailed
expressions to appendices \ref{App:Fcharge} and \ref{App:FII}. In
Sec.~\ref{sec:results}, we discuss the numerical results, and we
conclude in Sec.~\ref{sec:summary}.

\section{Theoretical Building Blocks}
\label{sec:theor-build-blocks}
The leading order CP-even effective Lagrangian in EFT($\slashed{\pi}$)
for the three-nucleon system is~\cite{Bedaque:1999ve}
\begin{eqnarray} 
\nonumber
  \mathcal{L}&=& N^{\dagger}\left(i \partial_{0} + e A_0 \frac{1 + \tau_3}{2} +\frac{\vec{\nabla}^{2}}{2 M_{N}}\right) N
  + \Delta_{t} t_{i}^{\dagger} t_i + \Delta_{s} s_a^\dagger s_a 
+y_t \left[ t_i^\dagger N^T \hat{P}_t^i N+\mathrm{H.c.}\right]\\
&&\hspace{-3mm}
  +y_{s}\left[ s_a^\dagger N^T \hat{P}_s^a N+\mathrm{H.c.}\right]
  +
  \Omega \psi^\dagger \psi
 + \left[\omega_{t}  \psi^\dagger \sigma_{i} N t_i +\text { H.c. }  \right] 
  - \left[ \omega_{s} \psi^\dagger \tau_{a} N s_a +\text { H.c. } \right] ~,
  \label{LOLag}
\end{eqnarray}
where the auxiliary dimer fields $t_i$ and $s_a$ represent the $^3S_1$
and $^1S_0$ dibaryon field, respectively.  The trimer field $\psi$
represents the three-nucleon field with total angular momentum
$1/2$. A three-nucleon force appears at LO because it was shown
\cite{Bedaque:1999ve, Bedaque:1998kg, Bedaque:1998km} to be necessary
for the renormalization of three-body observables.

The operators
$\hat{P}_t$ and $\hat{P}_s$, 
\begin{equation}
  \hat{P}_t^i =  \frac{1}{\sqrt{8}} \sigma^2 \sigma^i \tau^2,\qquad
  \hat{P}_s^a =  \frac{1}{\sqrt{8}} \sigma^2 \tau^2 \tau^a,
\end{equation}
project on the spin-triplet, isospin-singlet and spin-singlet,
isospin-triplet channels, respectively. For the coefficients in
Eq.~\eqref{LOLag} we choose the conventions
\begin{align}
  y_t^2 = y_s^2 = \frac{4\pi}{M_N},\quad
  \Delta_t = \gamma_t - \mu, \quad
  \Delta_s = \gamma_s - \mu, 
\end{align}
where $\gamma_t \simeq 45.7$ MeV denotes the binding momentum of the
deuteron, and $\gamma_s \simeq -7.9$ MeV is the $^1S_0$ virtual-state
momentum.  The renormalization scale $\mu$ is introduced through the
use of the so-called power divergence subtraction scheme in the
two-nucleon sector \cite{Kaplan:1998tg}.

Using a matching calculation to a theory without trimer fields it can
be shown that $\omega_s = \omega_t$~\cite{Vanasse:2015fph}. These
parameters are functions of the ultraviolet cutoff in the
three-nucleon Schr\"odinger equation. They are determined by adjusting
them (at a given cutoff) to a three-nucleon observable such as a
binding energy, e.g. $B(^3{\rm H}) = - 8.48$~MeV.

The dressed spin-triplet and spin-singlet dibaryon propagators are calculated
by summing over an infinite number of loop diagrams. At LO, they are 
given by
\begin{equation}
  iD^{\text{LO}}_{t,s} (p_0, p)= \frac{i}{\gamma_{t,s} - \sqrt{\frac{ p ^2}{4} -M_N p_0 - i\epsilon} }~.
\end{equation}
The renormalization of the deuteron wave function at LO is given by
the residue about the deuteron pole,
\begin{equation}
  Z_d^{\text{LO}} = \frac{2\gamma_t}{M_N}~.
\end{equation}

CPV from BSM physics can be systematically classified in the framework
of the Standard Model Effective Field Theory (SMEFT)
\cite{Buchmuller:1985jz,Grzadkowski:2010es}, where the SM is
complemented by the most general set of higher-dimensional operators,
expressed in terms of SM fields and invariant under the SM gauge
group.  The most important CPV operators arise at canonical
dimension-six, and are suppressed by two powers of $v/\Lambda_{X}$,
where $\Lambda_X$ is the BSM physics scale and $v=246$ GeV is the
Higgs vacuum expectation value.  For EDM studies, heavy SM degrees of
freedom can be integrated out, by matching the SMEFT onto an
$SU(3)_c \times U(1)_{\rm em}$ invariant EFT
\cite{Jenkins:2017jig,Jenkins:2017dyc,Dekens:2019ept}. Focusing on two
light quark flavors and on operators that are induced by SMEFT
operators at tree level, the dimension-six CPV Lagrangian relevant for
light nuclear EDMs includes one dimension-four operator, the QCD
$\bar\theta$ term, and nine dimension-six operators, the gluon
chromo-electric dipole moment, the $u$ and $d$ quark electric and
chromo-electric dipole moments, and four four-fermion operators. The
operator set can be easily extended to include strange quarks
\cite{Jenkins:2017jig,Jenkins:2017dyc,Dekens:2019ept,Mereghetti:2018oxv}.

At low-energy, these operators manifest in CP-violating interactions
between nucleons and photons.  In the single nucleon sector, the most
important CPV operators are the neutron and proton EDMs,
\begin{eqnarray}
  \label{nucleonEDM}
  \mathcal L_{N\gamma} &=& -  e N^{\dagger} \left(d_p \frac{1+\tau_3}{2} + d_n \frac{1-\tau_3}{2}  \right)  \left(S^\mu v^\nu - S^{\nu} v^\mu\right) N F_{\mu \nu}
                           \nonumber \\
&=& 
    e N^{\dagger} \left(d_p \frac{1+\tau_3}{2} + d_n \frac{1-\tau_3}{2}  \right)
    \boldsymbol{ \sigma} \cdot \mathbf{E} \, N ,  
\end{eqnarray}
where $v^{\mu} = (1, \mathbf{0})$ and $S^{\mu}= (0, \pmb{\sigma}/2)$
in the nucleon rest frame, and $\mathbf{E}$ denotes the electric
field.  For all quark-level sources of CPV one expects $d_n \sim d_p$
\cite{Pospelov:2005pr,deVries:2010ah}, but the calculation of the
exact dependence of $d_{p,n}$ on CPV quark-level couplings requires
non-perturbative techniques.  The momentum dependence of the nucleon
EDFF was computed in
Refs. \cite{Hockings:2005cn,deVries:2010ah}. Since the typical scale
of the momentum variation is $q\sim m_\pi$, we ignore it in this
paper.

For the QCD $\bar\theta$-term, the neutron EDM can be estimated by the
size of the long-range pion loop
\cite{Crewther:1979pi,Narison:2008jp,Hockings:2005cn,Ottnad:2009jw,deVries:2010ah,Mereghetti:2010kp,Seng:2014pba,deVries:2015una}
\begin{eqnarray}\label{nEDM}
d_n(\bar\theta) \simeq 2  \cdot 10^{-3} \, \bar\theta \, e \, \textrm{fm},
\end{eqnarray}
in good agreement with the naive expectation
$d_n = \mathcal O( m_\pi^2/\Lambda_\chi^3 \, \bar\theta)$, where
$\Lambda_\chi = 2\pi F_\pi$ is the chiral perturbation theory
breakdown scale, with $F_\pi \simeq 92$ MeV the pion decay constant.  Progress in lattice QCD calculations will soon allow
a theoretical error to be attached to the estimate in Eq. \eqref{nEDM}
\cite{Izubuchi:2017evl,Abramczyk:2017oxr,Bhattacharya:2018qat,Syritsyn:2019vvt,Kim:2018rce,Dragos:2019oxn}.
The nucleon EDM induced by dimension-six operators has been estimated
using QCD sum rules
\cite{Pospelov:1999ha,Pospelov:1999mv,Pospelov:2005pr,Haisch:2019bml}
or chiral techniques
\cite{deVries:2012ab,Seng:2014pba,Cirigliano:2016yhc}.  With the
exception of the contribution of the quark EDM, which is determined by
the nucleon tensor charges \cite{Gupta:2018lvp,Aoki:2019cca}, these
estimates have large uncertainties.

In EFT($\slashed{\pi}$), the leading two-nucleon operators resulting
in a non-zero EDM are given by
\begin{multline}
  \label{eq:2nucleon-L-edm}
     \mathcal{L}_{\slashed{P}\slashed{T}} =  \frac{y_t}{\sqrt{8}} C_{^3 S_1 - ^1P_1} \left( t_i^{\dagger} N^t \sigma_2 \tau_2 \overleftrightarrow \nabla^i N  \right) 
     + \frac{y_t}{\sqrt{8}} C_{^3 S_1 - ^3P_1} i \varepsilon^{il m} \left( t_i^{\dagger} N^t \sigma_2 \tau_2 \tau_3  \overleftrightarrow \nabla^m \sigma^l N  \right) \\
     + \frac{y_s}{\sqrt{8}}  \left( s_a^{\dagger} N^t \sigma_2 \tau_2 \tau_b \pmb{\sigma} \cdot \overleftrightarrow \nabla N  \right) 
     \left( C^{(0)}_{^1 S_0 - ^3P_0} \delta^{ab} + C^{(1)}_{^1 S_0 - ^3P_0} i \varepsilon^{3ab} +  C^{(2)}_{^1 S_0 - ^3P_0} \left(\delta^{ab} - 3 \delta^{a3} \delta^{b3}\right) \right)~.
\end{multline}
These operators were constructed in
Refs. \cite{Maekawa:2011vs,Vanasse:2019fzl,deVries:2020iea}. 
All operators mediate transitions between $S$ and $P$ waves, as denoted by the
name of the coefficients.  The operators $C_{^3 S_1 - ^1P_1}$ and
$C^{(0)}_{^1 S_0 - ^3P_0}$ are isospin invariant, $C_{^3 S_1 - ^3P_1}$
and $C^{(1)}_{^1 S_0 - ^3P_0}$ break isospin by one unit, while
$C^{(2)}_{^1 S_0 - ^3P_0}$ is an isotensor operator.  The couplings
$C_{^3 S_1 - ^1P_1}$, $C_{^3 S_1 - ^3P_1}$ and
$C^{(i)}_{^1 S_0 - ^3P_0}$ have dimension of $\textrm{mass}^{-1}$, and
are independent of the renormalization scale $\mu$.
Ref.~\cite{Maekawa:2011vs} provides a naive-dimensional-analysis
estimate of the size of these coefficients in terms of quark-level
couplings. 
For example, in the case of the QCD $\bar\theta$-term we expect only
isospin-invariant operators to appear at leading order, with the
scaling
\begin{eqnarray}
  \label{eq:theta_estimate}
  C_{^3 S_1 - ^1P_1}=\frac{m_\star \bar\theta}{\Lambda^2_{\slashed{\pi}}} c_{^3 S_1 - ^1P_1}~,\quad
  C^{(0)}_{^1S_0 - ^3P_0} = \frac{m_\star \bar\theta}{\Lambda^2_{\slashed{\pi}}} c^{(0)}_{^1S_0 - ^3P_0}~,
\end{eqnarray}
where $m^{-1}_\star = m_u^{-1} + m_d^{-1}$, $\Lambda_{\slashed{\pi}}$
denotes the breakdown scale of EFT$(\slashed{\pi})$ and
$ c_{^3 S_1 - ^1P_1}$ and $c^{(0)}_{^1S_0 - ^3P_0}$ are numbers of
order one. Going beyond dimensional analysis requires first principle
calculations of CPV matrix elements.

In this work we will thus express the EDMs of $^3$H and $^3$He in terms of
$d_n$, $d_p$ and of the five couplings in
Eq. \eqref{eq:2nucleon-L-edm}, and discuss the minimal set of
observables that is necessary to disentangle them.

\subsection{The three-nucleon bound state vertex function}\label{sec:Gfunc}
We will calculate the EDFF  by integrating over
three-particle irreducible diagrams that contain a single insertion of
a CPV operator. Following the formalism defined in Refs. \cite{Vanasse:2015fph,Vanasse:2017kgh},
we define a diagram to be three-particle
irreducible when it cannot be separated by cutting at a trimer field
vertex.  The resulting form factor diagrams contain necessarily
infinite sums of nucleon-deuteron rescattering diagrams that are given
by vertex functions that result from an integral equation, and
pieces that include the photon coupling to a single nucleon line.
\begin{figure}
  \includegraphics[width=0.9\textwidth]{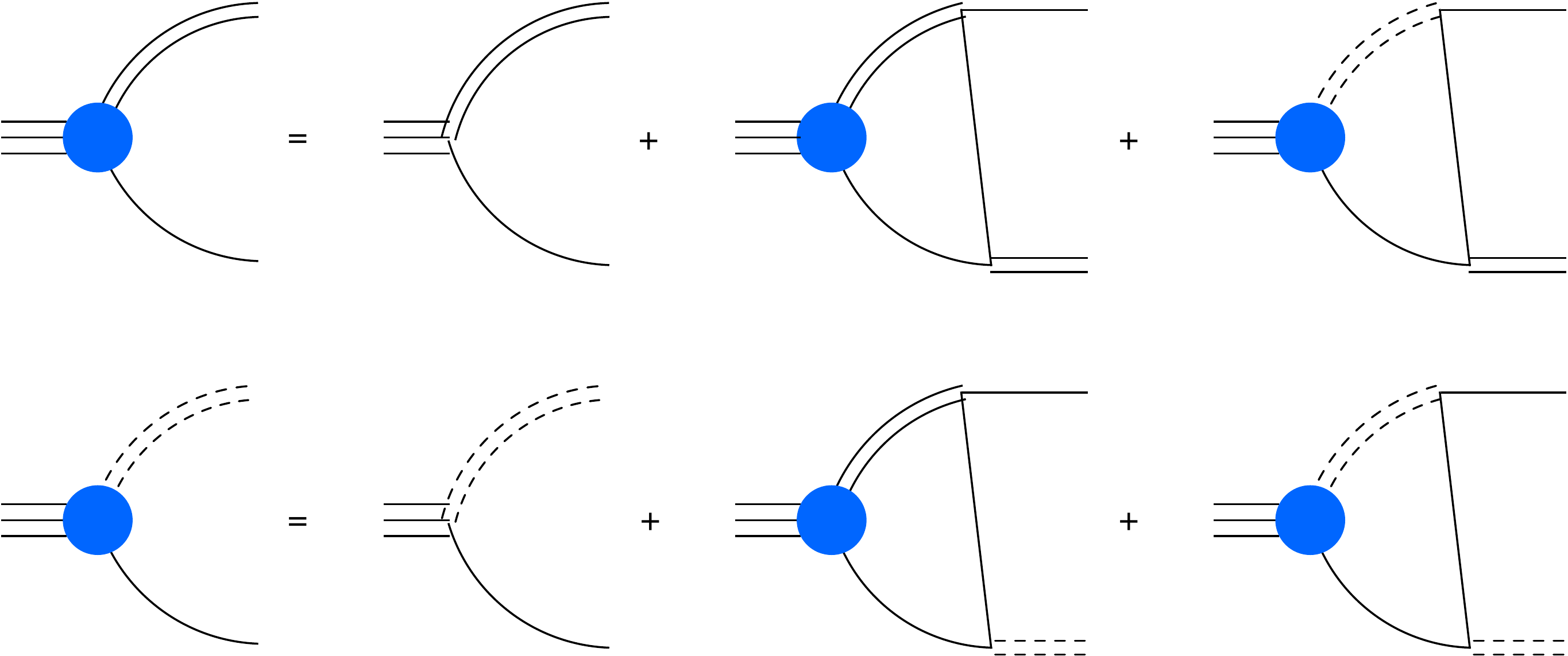}
  \caption{\label{fig:vertex-function-CP-even} Diagrammatic
    representation of the LO three-body CP-even vertex function. The
    (dashed) double line denotes a dressed spin-singlet (spin-triplet)
    dibaryon propagator.}
\end{figure}

The LO vertex function $\bm{\mathcal{G}}(E, p)$ [[for a three-nucleon
system in the center-of-mass frame with binding energy $E$ and relative
momentum ${\bf p}$ between outgoing nucleon and dimer]] is given
by the integral equation shown diagrammatically in
Fig.~\ref{fig:vertex-function-CP-even} and given explicitly by
\begin{equation}
  \label{eq:CPeven-vertex}
  \bm{ \mathcal{G} } (E, p) = \widetilde{\mathbf{1}} +\mathbf{K}^{0}(q, p, E) \otimes_q  \bm{  \mathcal{ \widetilde{G} } } (E, q)~.
\end{equation}
We define the short-hand notation
\begin{equation}
  \bm{  \mathcal{ \widetilde{G} } } (E, q) = 
  \bm{ D } \left(E-\frac{q^{2}}{2 M_{N}}, q\right)
  \bm{ \mathcal{ G } }(E, q)
\end{equation}
where  
\begin{equation}
  \bm{ D } \left(E-\frac{q^{2}}{2 M_{N}}, q\right) = 
  \left( \begin{matrix}
    D_t\left(E-\frac{q^{2}}{2 M_{N}}, q\right) & 0 \\
    0 & D_s\left(E-\frac{q^{2}}{2 M_{N}}, q\right) 
  \end{matrix}\right),
\end{equation}
and the inhomogeneous term in this integral equation is
\begin{equation}
  \widetilde{\mathbf{1}} =\left( \begin{array}{rr} 1  \\ -1 \end{array}\right)~.
\end{equation}
The convolution operator $\otimes_q$ is defined as
\begin{equation}
  A(q) \otimes_q B(q) = \int_0^\Lambda dq  \frac{q^2}{2\pi^2} A(q) B(q)~,
\end{equation}
where $\Lambda$ is a hard momentum-space cutoff. Observables will be $\Lambda$-independent for large cutoffs.

The homogeneous term is defined by
\begin{equation}
  \mathbf{K}^{\ell}(q, p, E)= R_\ell (q, p, E)  \left(\begin{array}{rr} -1 & 3 \\ 3 & -1\end{array}\right)~,
\end{equation}
where the function $R_\ell$ is defined as 
\begin{align}
  R_\ell (q, p, E) &= \frac{2 \pi}{q p} Q_{\ell}\left(\frac{q^{2}+p^{2}-M_{N} E-i \epsilon}{q p}\right) ~,
\end{align}
and $Q_l$ are functions proportionial to Legendre function of the
second kind but differ from their conventional definition by a phase of $(-1)^\ell$,
\begin{equation}
  Q_{\ell}(a)=\frac{1}{2} \int_{-1}^{1} \frac{P_{\ell}(x)}{a+x} \hbox{d}x~.
\end{equation}

\section{The T-odd vertex function}\label{ToddV}

\begin{figure}
     \includegraphics[width=\textwidth]{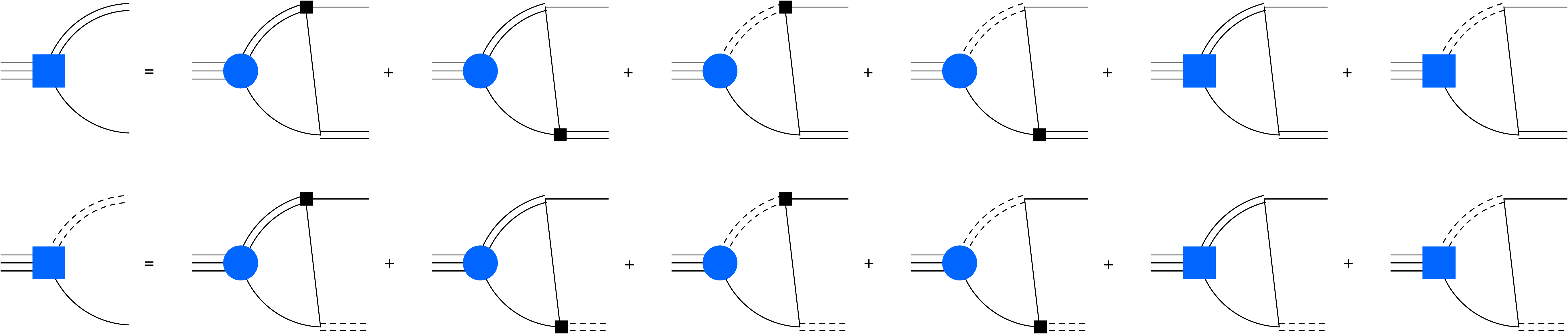}
     \caption{\label{fig:vertex-function-cp-odd} $\slashed{P}\slashed{T}$ vertex function. A blue square denotes
     the $\slashed{P}\slashed{T}$ vertex function, while a black square an insertion of the 
     operators in Eq. \eqref{eq:2nucleon-L-edm}. Remaining notation as in Fig. \ref{fig:vertex-function-CP-even}.
     }
\end{figure}
A three-particle irreducible diagram can contain repeated
nucleon-dimer scattering between a nucleon-photon vertex and an
insertion of a two-nucleon CP-odd vertex. We include these diagrams
through two integral equations that generate a vertex function that
contains a single insertion of the CP-odd two-nucleon interaction. The
diagrammatic expression for these vertex functions is shown in
Fig.~\ref{fig:vertex-function-cp-odd}.

The T-odd vertices convert the spin $1/2$, isospin $1/2$ trimer field
$\psi$ into a nucleon-dimer with three possible spin/isospin quantum
numbers: spin and isospin $1/2$, spin $1/2$ and isospin $3/2$, as well as spin
$3/2$ and isospin $1/2$.  The latter does not contribute to the
three-nucleon EDM at leading order, since the LO electromagnetic
interaction does not change spin and the overlap of the spin 3/2 T-odd
function with the triton or helion vanishes.  The integral equations for the
isospin 1/2 component, $\bm{ \mathcal{T} }^{ \frac{1}{2} }$, and the
isospin 3/2 component, $\bm{ \mathcal{T} }^{ \frac{3}{2} }$, of the spin-1/2
T-odd vertex functions are given by 
\begin{align}
  &(\pmb{\sigma} \cdot {\bf k}) 
   \bm{ \mathcal{T} }^{ \frac{1}{2} } (E, k) 
  =     (\pmb{\sigma} \cdot {\bf k})    \left\lbrace \bm{R}^{ \frac{1}{2} }_{\mathcal{T}} (E, k,q)    \otimes_q  
  \widetilde{ \bm{ \mathcal{T} } }^{\frac{1}{2}} (E,q) 
  + \bm{R}^\frac{1}{2} (E, k,q)    \otimes_q  \bm{ \widetilde{\mathcal{G}} } (E,q)
  \right\rbrace~,\\ 
  \nonumber
  &(\pmb{\sigma} \cdot {\bf k})  (\delta^{3c}+\tau^3\tau^c) 
   \bm{ \mathcal{T} }^{ \frac{3}{2} } (E, k) 
  = (\pmb{\sigma} \cdot {\bf k}) (\delta^{3c}+\tau^3\tau^c)
      \Big\lbrace \bm{R}^{ \frac{3}{2} }_{\mathcal{T}} (E, k,q)   
      \otimes_q  
  \widetilde{ \bm{ \mathcal{T} } }^{\frac{3}{2}} (E,q) 
  \\
  &
    \qquad\qquad\qquad\qquad
    \qquad\qquad\qquad+ \bm{R}^\frac{3}{2}(E, k,q)   
   \otimes_q  \bm{ \widetilde{\mathcal{G}} } (E,q)
  \Big\rbrace~,
\end{align}
where we show explicitly the spin/isospin structure of the vertex
functions, and, similarly to the CP-even case, we introduced the
shorthand notation for the product of a vertex function and a dressed
dibaryon propagator
\begin{equation}
  \widetilde{ \bm{ \mathcal{T} } }^{\frac{1}{2},\frac{3}{2}} (E, q) 
  = \bm{D} \left(E - \frac{q^2}{2 M_N}, q \right) 
  \bm{ \mathcal{T} }^{\frac{1}{2},\frac{3}{2}}  (E, q)~.
\end{equation}
The kernels of the homogeneous terms are  
\begin{align}
  \bm{ R }^{ \frac{1}{2} }_{\mathcal{T}}(E, k,q)   
  &= \frac{q}{k} R_1(E, k,q)   
  \begin{pmatrix}
    -1 & \phantom{-}3 \\
    \phantom{-}3 & -1
  \end{pmatrix} ,
  \qquad 
  \bm{ R }^{ \frac{3}{2} }_{\mathcal{T}} (E, k,q)   
  = \frac{q}{k} R_1(E, k,q) 
  \left( \begin{matrix}
    0 & 0 \\
    0 & 2
  \end{matrix} \right).
\end{align}
The inhomogeneous terms are driven by the T-odd operators in
Eq.~\eqref{eq:2nucleon-L-edm}.  The isospin 1/2 vertex functions
receive contributions from both isoscalar and isovector operators in
Eq.~\eqref{eq:2nucleon-L-edm}, 

\begin{align}
  \bm{ R }^{ \frac{1}{2} } (E, k,q)   
  & =    \left[
 R_0(E, k,q) \begin{pmatrix}
    -1 & 1 \\
    -2 & 0
  \end{pmatrix}
  + \frac{q}{k} R_1(E, k,q) 
  \left( \begin{matrix}
    \phantom{-}1 & 2 \\
    -1 & 0
  \end{matrix} \right) \right]
     \left( C_{^3S_1 - ^1P_1} + \frac{2}{3} \tau^3 C_{^3S_1 - ^3 P_1} \right)   \nonumber  \\
   & 
+\left[  R_0(E, k,q) \left( \begin{matrix}
  \phantom{+}0 & 2 \\
    -1 & 1
  \end{matrix} \right) - \frac{q}{k} R_1(E, k,q)  \left( \begin{matrix}
    0 & -1 \\
    2 & \phantom{+}1
  \end{matrix} \right)  \right]
  \left( C^{(0)}_{^1S_0 - ^3P_0} - \frac{2}{3} \tau^3 C^{(1)}_{^1S_0 - ^3 P_0} \right).
\end{align}

The isospin $3/2$ component is induced by the isotensor operator
$C^{(2)}_{^1S_0 - ^3 P_0}$ and by the isovector operators yielding

\begin{eqnarray}
  \bm{ R }^{ \frac{3}{2} } (E, k,q)
 & = & \left[ 2 R_0(E, k,q)  + \frac{q}{k} R_1(E, k,q) \right] \left(\begin{matrix}
 0 & 0 \\
 1 & 0
 \end{matrix}\right)  \frac{4}{3} C^{}_{^3S_1 - ^3P_1} \nonumber \\
 & & - \frac{1}{3}\left[  R_0(E, k,q) \left(\begin{matrix}
 0 & 0 \\
 1 & 5
 \end{matrix}\right)  + \frac{q}{k} R_1(E, k,q) \left(\begin{matrix}
 0 & 0 \\
 2 & 4
 \end{matrix}\right) \right]\left(C^{(1)}_{^1S_0 - ^3 P_0} - 3 \tau_3 C^{(2)}_{^1S_0 - ^3 P_0} \right).\nonumber\\
\end{eqnarray}

\subsection{Integral equations in the SU(4) limit}\label{ToddV_SU4}
Nuclear interactions exhibit an approximate $SU(4)$ spin-isospin
(Wigner) symmetry, which would be exact in the 
limit~\cite{PhysRev.51.106, Vanasse:2016umz} of equal spin-triplet
and singlet scattering lengths.  $SU(4)$ breaking is
parameterized by the difference $\gamma_t - \gamma_s$, and the
expansion around the Wigner limit converges very well
\cite{Vanasse:2016umz}.  We will study the electric dipole form factor
of the three-nucleon system in the $SU(4)$ limit, and provide the
relevant formulae in this section.

In the $SU(4)$ limit, $D_t = D_s = D_{SU(4)}$ and from
Eq.~\eqref{eq:CPeven-vertex} one can see that
$\widetilde{\mathcal G}_t = - \widetilde{\mathcal G}_s$.  We can
introduce the combinations
\begin{equation}\label{su4_g}
{\mathcal G}_{\pm} = \frac{1}{2} \left( {\mathcal G}_t \mp {\mathcal G}_s \right),
\end{equation}
so that $\widetilde{\mathcal G}_{-}$ vanishes in the $SU(4)$ limit. 

The structure of the T-odd vertex functions simplifies significantly in the $SU(4)$ limit. It can be shown
that both the isospin 1/2 and isospin 3/2 components  are proportional to a single 
function $\mathcal T_{SU(4)}$, which satisfies the integral equation
\begin{align}
  \label{su4_7}
  \mathcal T_{SU(4)}(E,k) &=  - \mathcal R_{SU(4)}(E,k,q) \otimes_q \widetilde{\mathcal G}_+(E,q) + 2\frac{q}{k} R_1(E,k,q)  \otimes_q    \widetilde{\mathcal T}_{SU(4)}(E,q)~, \\
 \mathcal R_{SU(4)}(E,k,q) &= 2 R_0(E, k, q) + \frac{q}{k}  R_{1}(E, k, q).
\end{align}

In terms of $\mathcal T_{SU(4)}$, we can write
\begin{eqnarray}
  \bm{ \mathcal{T} }_{SU(4)}^{\frac{1}{2}}(E, k) 
  = \left( \begin{matrix}
    1 \\ 1
  \end{matrix}\right)
  \mathcal{T}_{SU(4)}^{\frac{1}{2}}(E, k)~,\quad
  \bm{ \mathcal{T} }_{SU(4)}^{\frac{3}{2}}(E, k) 
  = \left( \begin{matrix}
    0 \\ 1
  \end{matrix}\right)
  \mathcal{T}_{SU(4)}^{\frac{3}{2}}(E, k)~,
\end{eqnarray}
where
\begin{align}
  \mathcal{T}_{SU(4)}^{\frac{1}{2}}(E, k) & =  
  \bigg[   C^{(0)}_{^1S_0 - ^3P_0}  + C^{}_{^3S_1 - ^1P_1} + 
  \frac{2\tau^3}{3} ( C^{}_{^3S_1 - ^3P_1}    -  C^{(1)}_{^1S_0 - ^3P_0} )  \bigg]    \mathcal T_{SU(4)}(E,k)~, \\
  \mathcal{T}_{SU(4)}^{\frac{3}{2}}(E, k) & =  
  \bigg[  -\frac{2}{3}  ( 2 C^{}_{^3S_1 - ^3P_1}  + C^{(1)}_{^1S_0 - ^3P_0})
  + 2\tau^3  C^{(2)}_{^1S_0 - ^3P_0} \bigg]    \mathcal T_{SU(4)}(E,k)~.
\end{align}

\section{Three-nucleon form factors}
\label{sec:three-nucleon-form}

The EDFF of a three-nucleon system can be obtained from the matrix
element of the zero-component of the electromagnetic current $J^0$ in
the presence of CP violation. Neglecting recoil corrections, we can
write the matrix element of $J^0$ as
 \begin{equation}
   \label{eq:ME-J0}
  \langle {\bf p}', \alpha| J^0|{\bf p}, \beta \rangle
  = F_C(q^2) \delta_{\alpha \beta} +  \left[\pmb{\sigma}\cdot {\bf q}\right]_{\alpha\beta} F_D(q^2), 
\end{equation}
where $\alpha$ and $\beta$ are spin indices of the in- and outgoing
three-nucleon state, ${\bf q} = {\bf p} - {\bf p}'$ is the
momentum injected by the current, and $q = |\textbf{q}|$.  $F_C$ denotes the charge form
factor and $F_D$ the electric dipole form factor, which vanishes in
the absence of CP-violation.  We will write the EDFF in terms of two
components,
\begin{eqnarray}
F_D(q^2) = F_{\rm I}(q^2) + F_{\rm II}(q^2).
\end{eqnarray}
$F_{\rm I}$ denotes the EDFF generated by the T-odd component of the
electromagnetic current, which is dominated by one-nucleon operators,
namely the neutron and proton EDMs in Eqs.~\eqref{nucleonEDM}.  CPV
interactions can in addition generate a CP-odd component in the
three-nucleon wavefunction, which is dominated by the two-body
operators in Eq.~\eqref{eq:2nucleon-L-edm}. We denote the ensuing EDFF
by $F_{\rm II}$.

\begin{figure} 
\subfigure[]{
    \includegraphics[scale=0.5]{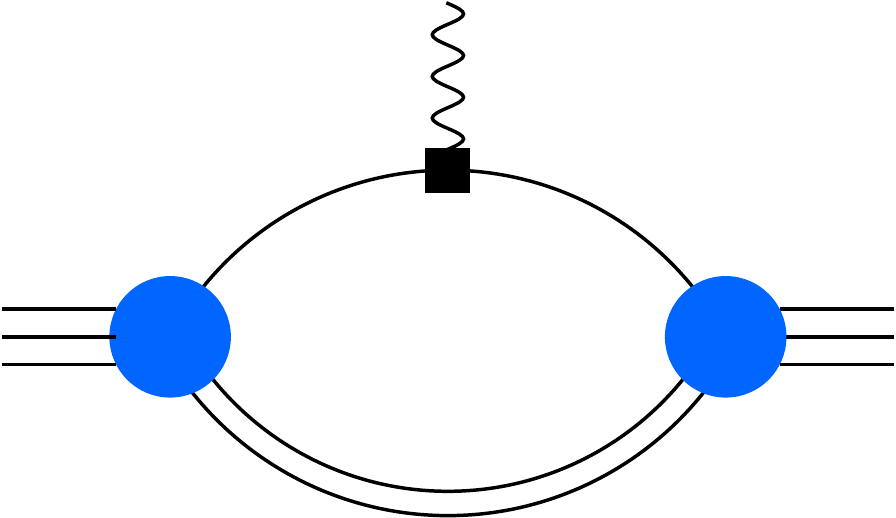}}
\subfigure[]{
    \includegraphics[scale=0.5]{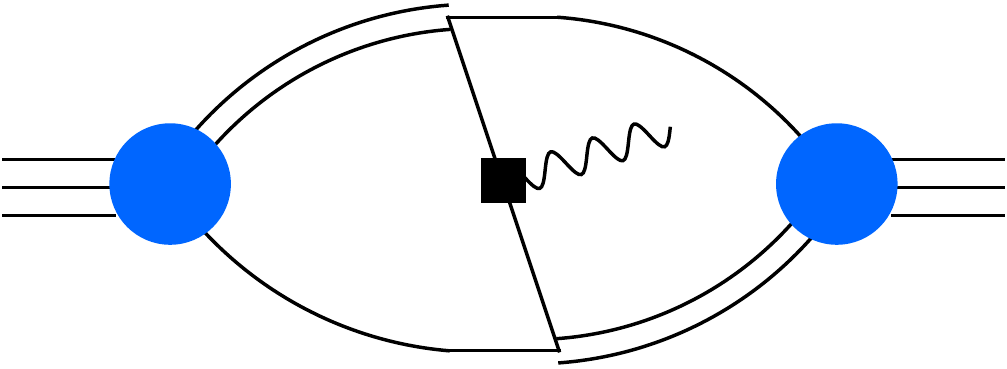}}
\subfigure[]{
    \includegraphics[scale=0.5]{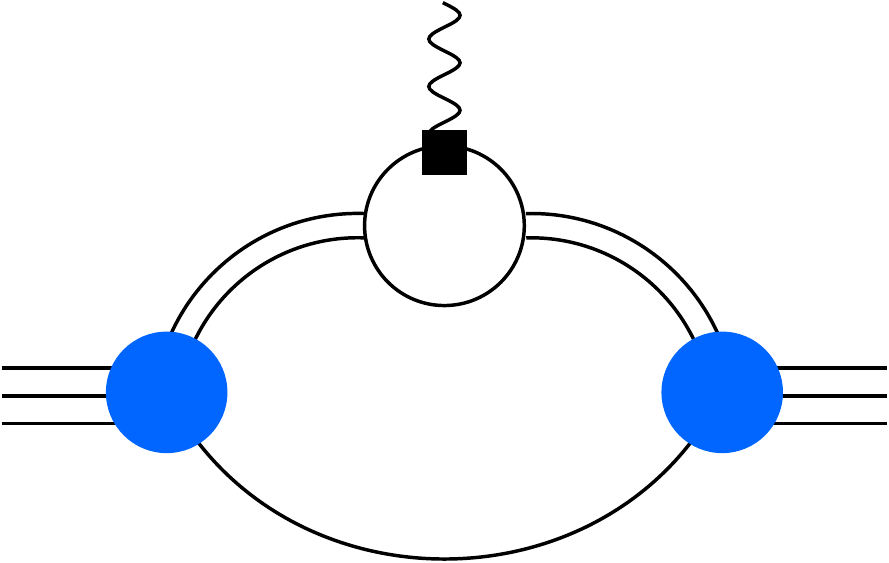}}
    \caption{Diagrams contributing to the one-body component of the
    EDFF, defined in Eq.~\eqref{eq:edff-one-body}. Here, the double
    line can denote a spin-triplet or singlet dimer. The black square
    denotes an insertion of the nucleon EDM, defined in
    Eq.~\eqref{nucleonEDM}.}
    \label{fig:one-body-diagrams}
\end{figure}

The diagrams contributing to $F_I$ are shown in 
Fig.~\ref{fig:one-body-diagrams}, where the black square 
denotes an insertion of the nucleon EDM, defined in Eq.~\eqref{nucleonEDM}.
We therefore write the $F_{\rm I}$ as the sum of
the three terms
\begin{equation} \label{eq:edff-one-body}
F_{\rm I}(q^2) =  F_{\rm I}^A(q^2) + F_{\rm I}^B(q^2) + F_{\rm I}^C(q^2) ~,
\end{equation}
corresponding to the three diagrams shown in
Fig.~\ref{fig:one-body-diagrams}.  We give explicit expressions for
the diagrams in Appendix \ref{App:Fcharge}.  From the expression in
Appendix \ref{App:Fcharge} and the charge form factor in
Refs.~\cite{Vanasse:2015fph,Vanasse:2017kgh}, which we also report in
Appendix \ref{App:Fcharge}, it can be seen that in the $SU(4)$ limit,
the one-body contribution to the triton and $^3$He EDFFs is identical
to $F_C(q^2)$, weighted by the proton or
neutron EDM, 
\begin{eqnarray}
  \label{eq:edff-one-body4}
F_{\rm I}(q^2, ^3{\rm H}) \xrightarrow{SU(4)}  d_p \, F_C(q^2), \qquad F_{\rm I}(q^2, ^3 {\rm He}) \xrightarrow{SU(4)} d_n  F_C(q^2).  
\end{eqnarray}
We will see that the results at the physical values of $\gamma_s$ and
$\gamma_t$ deviate from this expectation by a few percent.

\begin{figure}
\subfigure[]{\includegraphics[scale = 0.5]{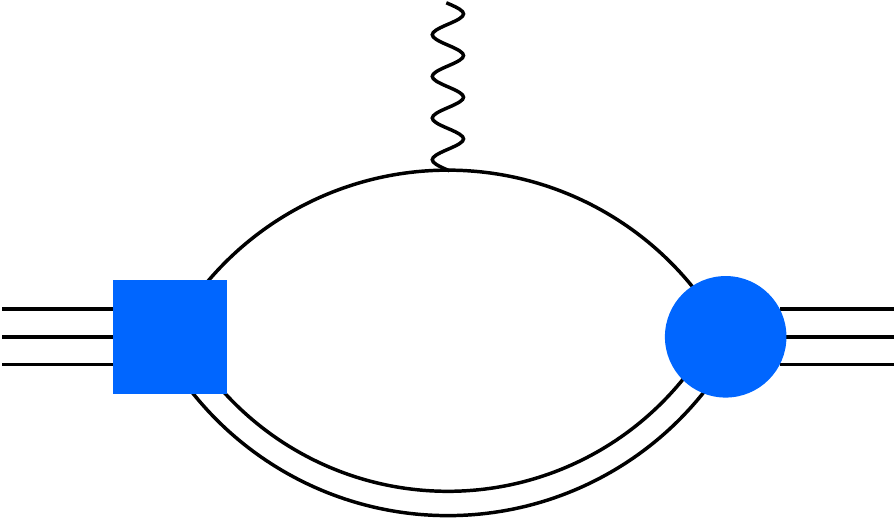}}
\subfigure[]{\includegraphics[scale = 0.5]{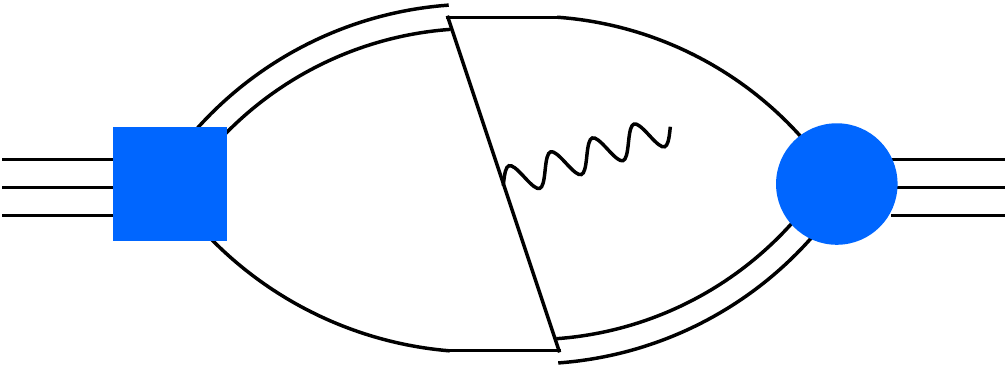}}
\subfigure[]{\includegraphics[scale = 0.5]{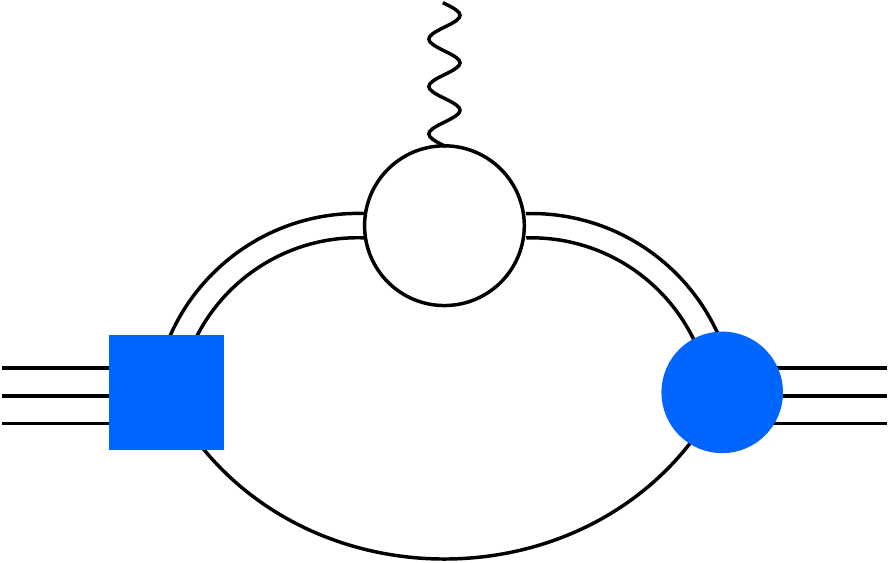}}
  \subfigure[]{
    \includegraphics[scale =0.5]{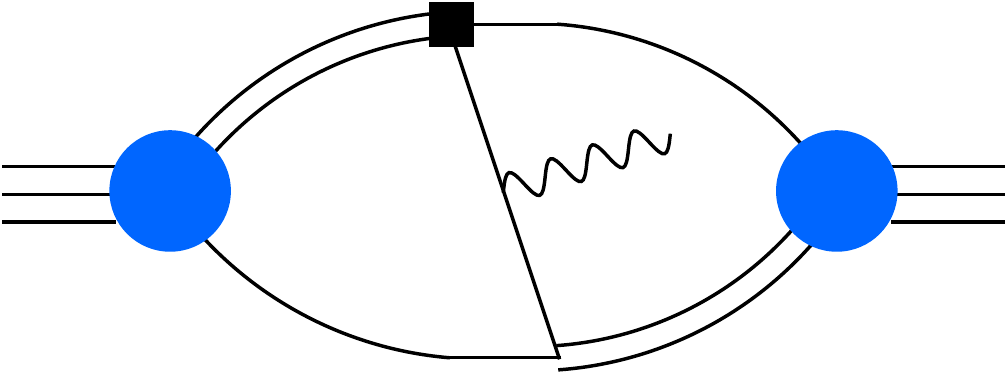}
    }
\subfigure[]{
    \includegraphics[scale = 0.5]{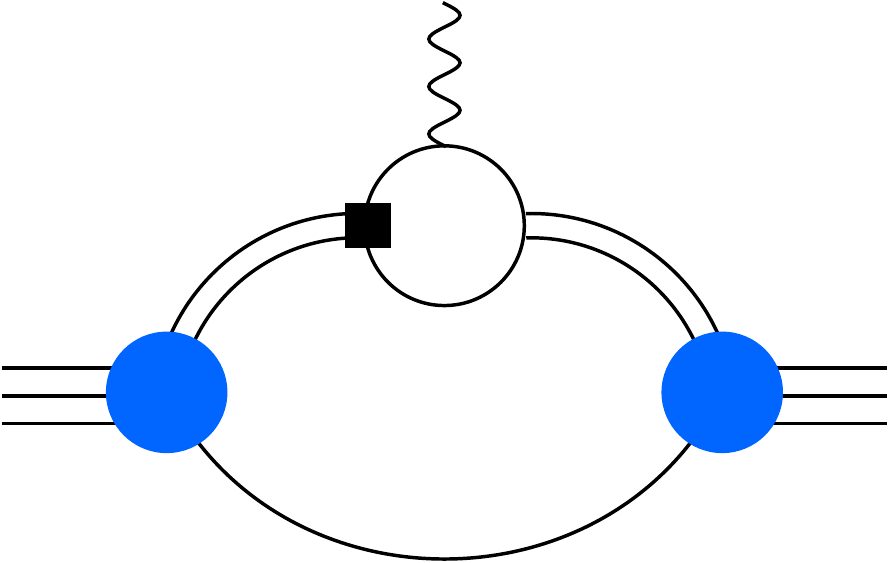}}
  \caption{Diagrams for the three-nucleon EDM form factor at LO that
    involve a CP-odd two-nucleon interaction.}
  \label{fig:two-body-diagrams}
\end{figure}
The second class of contributions arises from the two-nucleon
operators given in Eq.~\eqref{eq:2nucleon-L-edm}.  In
Fig.~\ref{fig:two-body-diagrams} we show the EDFF
topologies that include a CP-odd two-nucleon operator. Diagrams $(a)$,
$(b)$ and $(c)$ include the T-odd vertex functions defined in Section
\ref{ToddV}.  Diagram $(d)$ and $(e)$ include the CP-even $\mathcal G$
vertex functions, with an additional insertion of the T-odd
nucleon-dimer operators.  For simplicity, we only show one
topology. The complete set of diagrams also includes the insertions of
the T-odd operators to the left of the photon-nucleon vertex.

We write the sum of contributions to the EDFF that include
two-body CP-odd interactions as
\begin{equation}
  F_{\rm II}(q^2) = 
   F_{\rm II}^A(q^2) + F_{\rm II}^B(q^2) + F_{\rm II}^C(q^2)
  + F_{\rm II}^D(q^2) + F_{\rm II}^E(q^2) ~,
\end{equation}
where the superscript indicates the corresponding diagram in
Fig.\ref{fig:two-body-diagrams}. We give explicit expressions for the
individual diagrams in Appendix \ref{App:FII}.

In the $SU(4)$ limit, the two-body diagrams also undergo a noticeable
simplification, and they become proportional to a single combination of
T-odd coefficients,
\begin{eqnarray}
  \label{eq:edff-2body}
F_{\rm II}(q^2, ^3{\rm H})  &\xrightarrow{SU(4)}&  \tilde{F}_{SU(4)}(q^2) \left(  C_{^3S_1 - ^1P_1} + C^{(0)}_{^1S_0 - 3P_0} - 2 C^{(2)}_{^1S_0 - ^3 P_0} - 2 C_{^3S_1 - ^3 P_1} \right), \\
  F_{\rm II}(q^2, ^3{\rm He}) &\xrightarrow{SU(4)}&  -\tilde{F}_{SU(4)}(q^2) \left(  C_{^3S_1 - ^1P_1} + C^{(0)}_{^1S_0 - 3P_0} - 2 C^{(2)}_{^1S_0 - ^3 P_0} + 2 C_{^3S_1 - ^3 P_1} \right)~,
\end{eqnarray}
where $\tilde{F}_{SU(4)}(q^2)$ is a universal function that depends on $q$, on
the scattering length in the Wigner limit and the three-body binding energy.
In particular, the three-nucleon EDM becomes insensitive to the
isospin-1 $C^{(1)}_{^1S_0 - ^3P_0}$ operator.

\section{Results}
\label{sec:results}
We have calculated the numerical coefficients multiplying the
low-energy constants that appear in a decomposition of the CP-odd form
factor as a function of $q^2$. In the absence of the Coulomb
interaction, we take the binding energy of $^{3}$H and $^3$He to be
equal, {\it i.e.} $B(^3{\rm H}) = B(^3{\rm He})$. We estimate the
numerical uncertainty of the results presented below to be 1~\% or
lower. The theoretical uncertainty of our results is determined by the
expansion parameter of the pionless EFT which is
$\gamma_t \rho_t \approx 0.4$, where $\rho_t$ is the effective range
in the triplet channel. The theoretical uncertainties of our results
are therefore clearly larger than the numerical ones.

\begin{figure}[t]
  \subfigure[$C_{^3S_1-^1P_1}$]{
  \includegraphics[width = 0.4\textwidth]{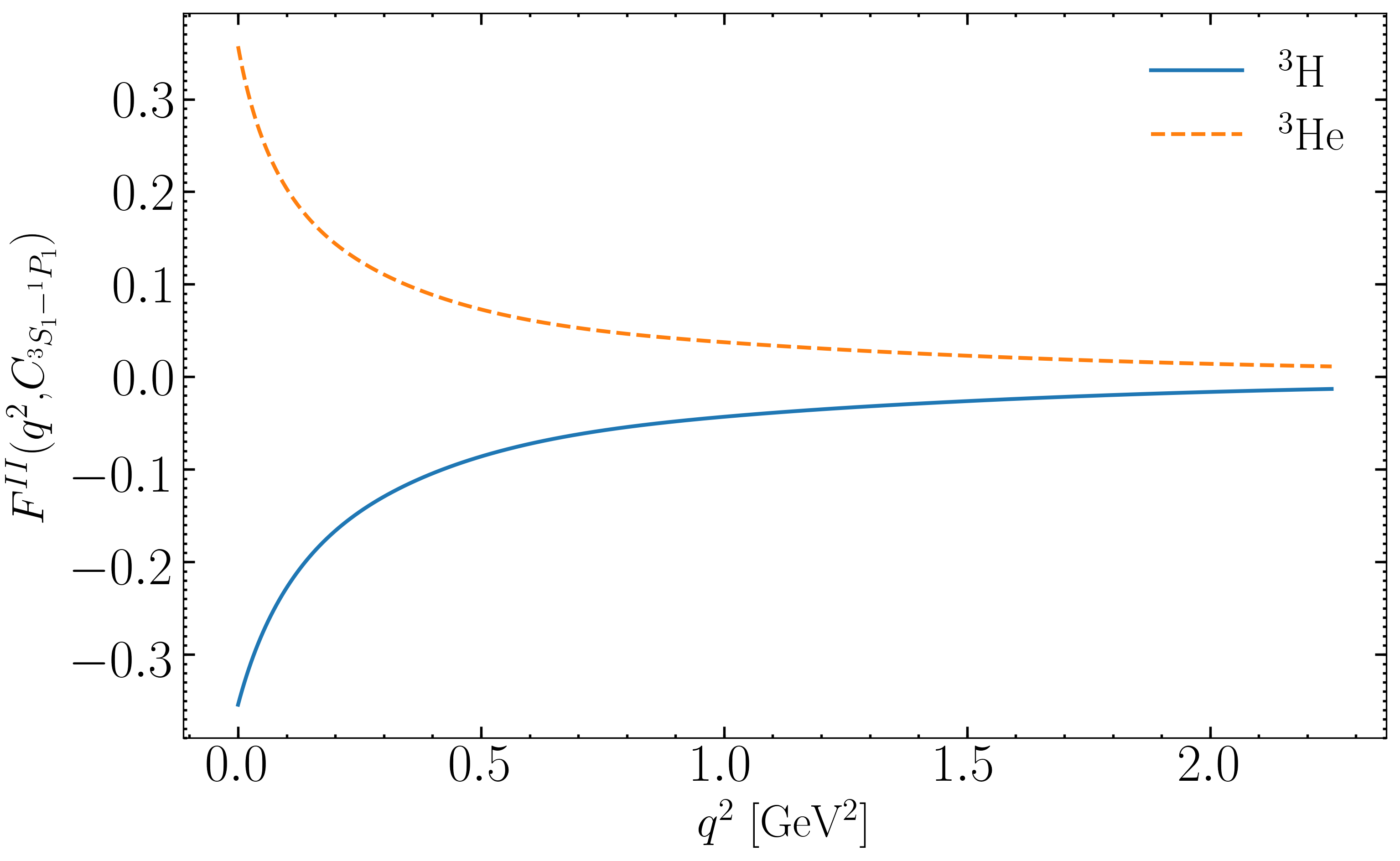}}
\subfigure[$C_{^3S_1-^3P_1}$]{
  \includegraphics[width = 0.4\textwidth]{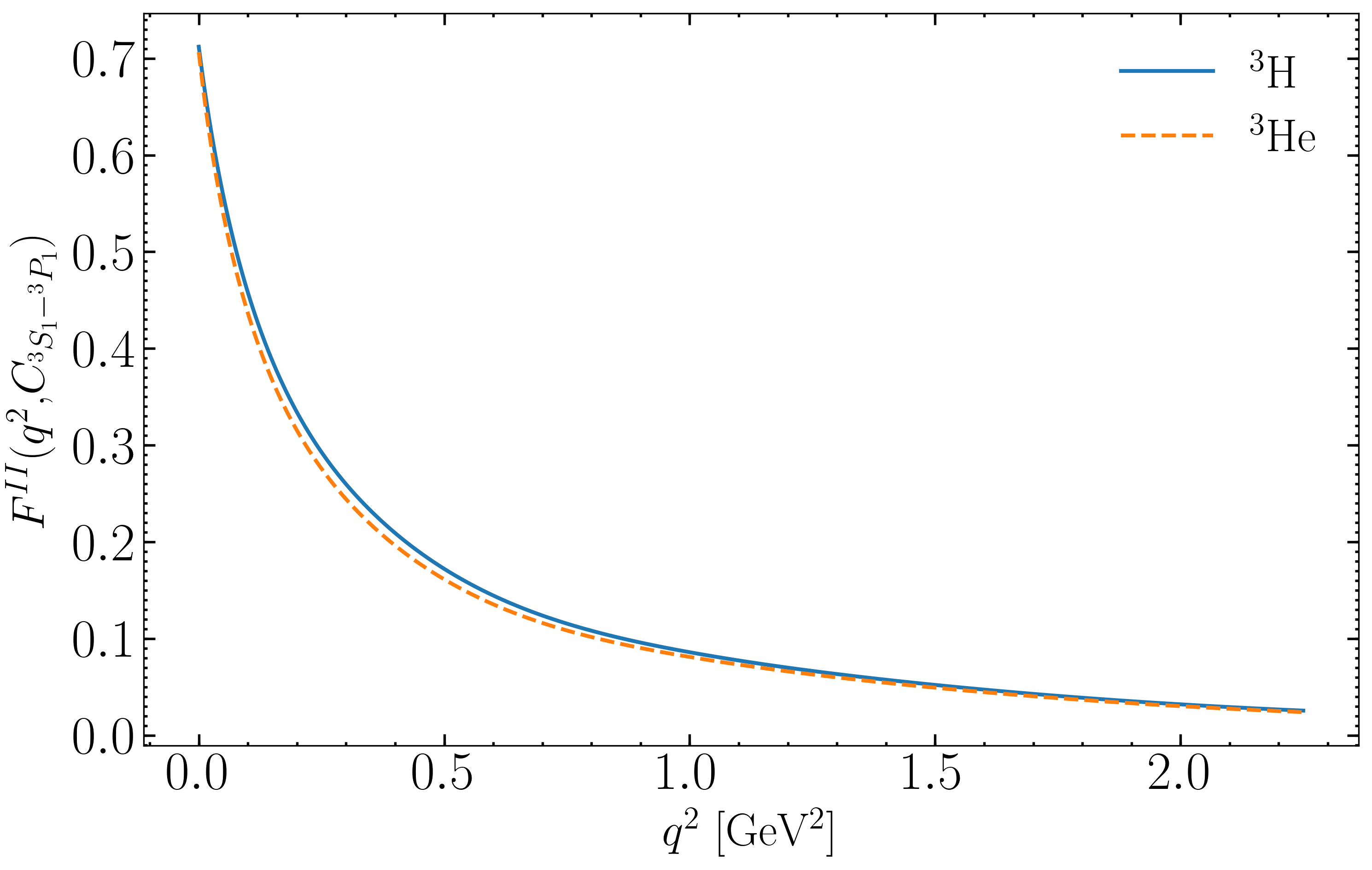}
}
\subfigure[$C_{^1S_0-^3P_0}^{(0)}$]{
  \includegraphics[width = 0.4\textwidth]{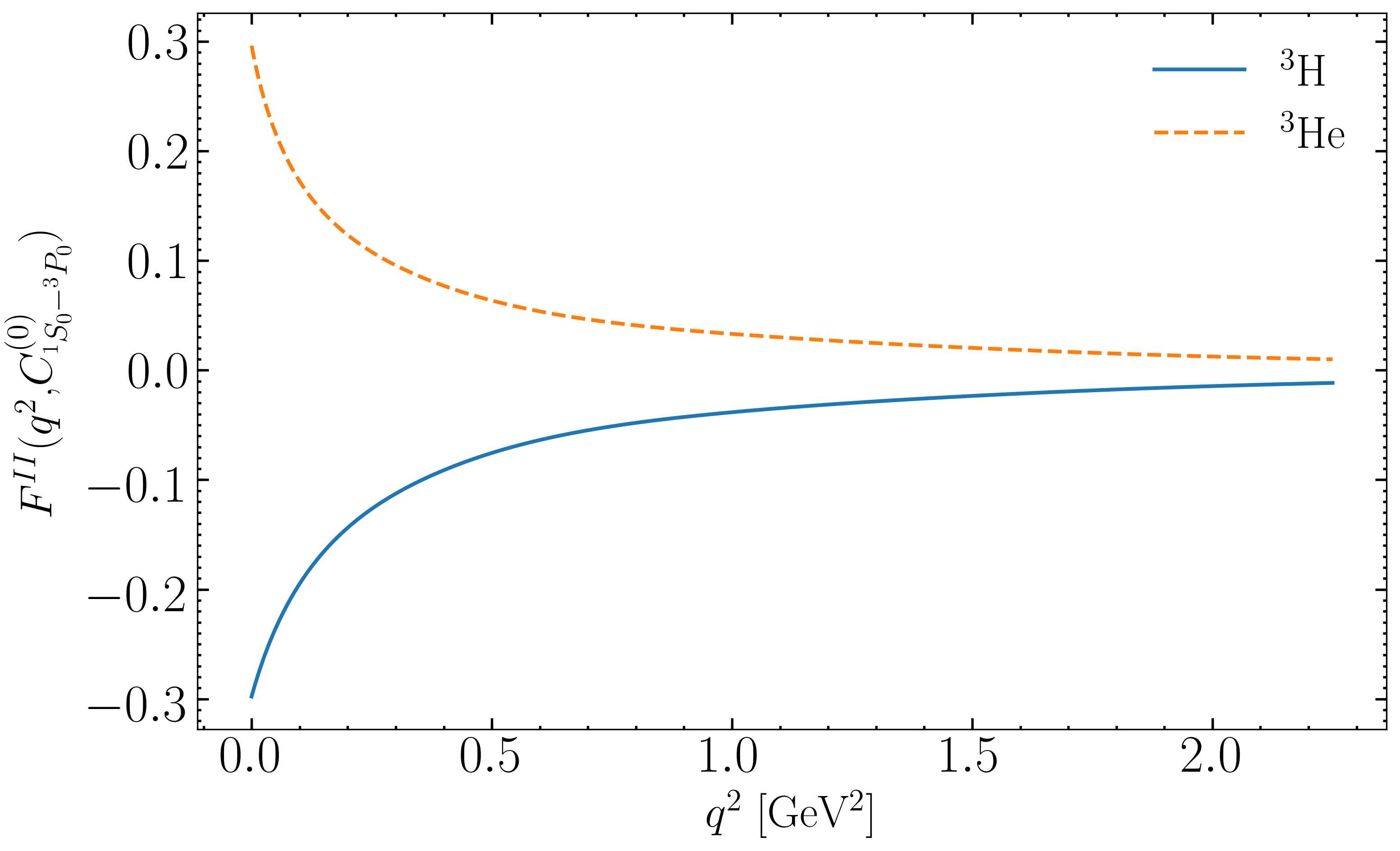}
}
\subfigure[$C_{^1S_0-^3P_0}^{(1)}$]{
  \includegraphics[width = 0.4\textwidth]{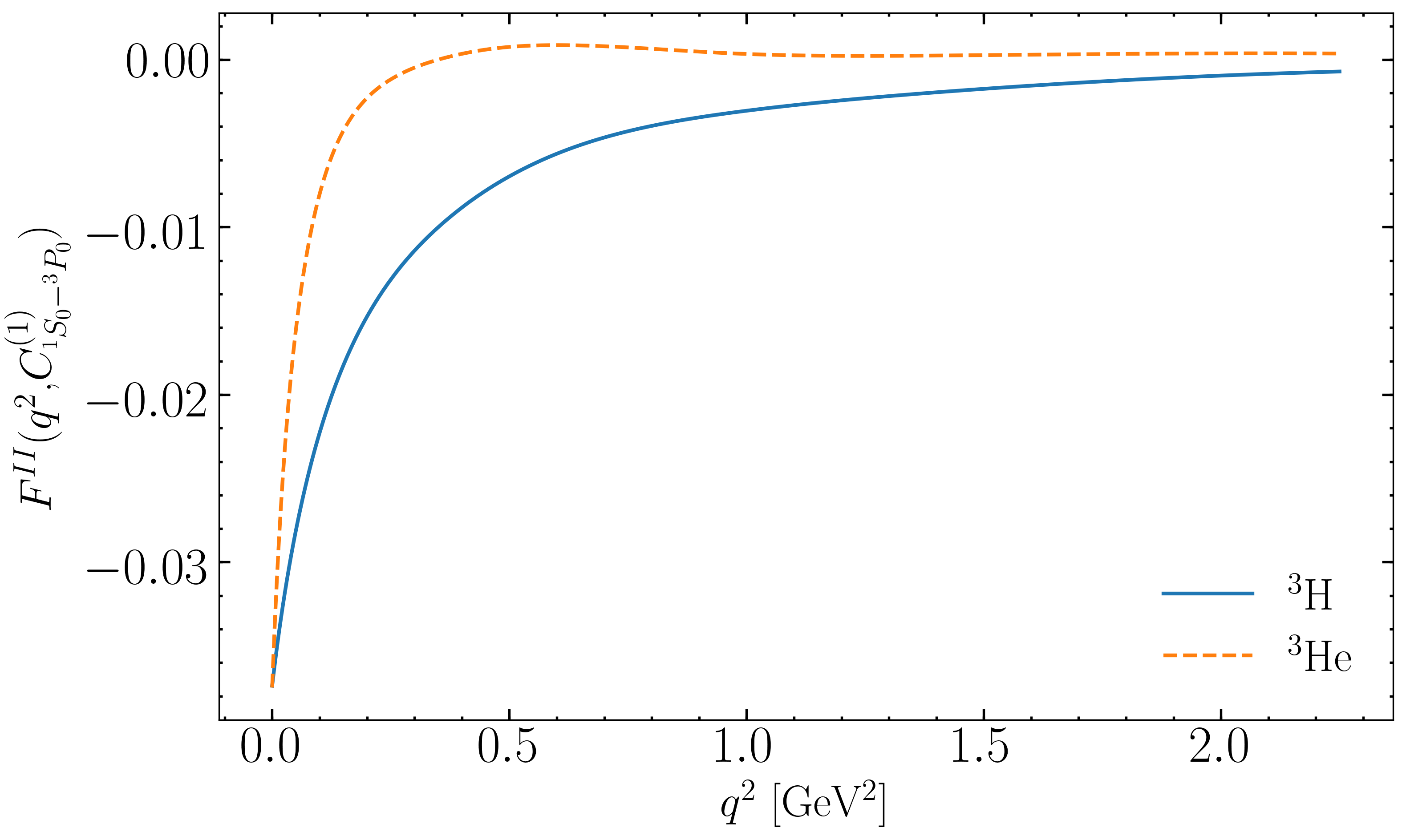}}
\subfigure[$C_{^1S_0-^3P_0}^{(2)}$]{
     \includegraphics[width = 0.4\textwidth]{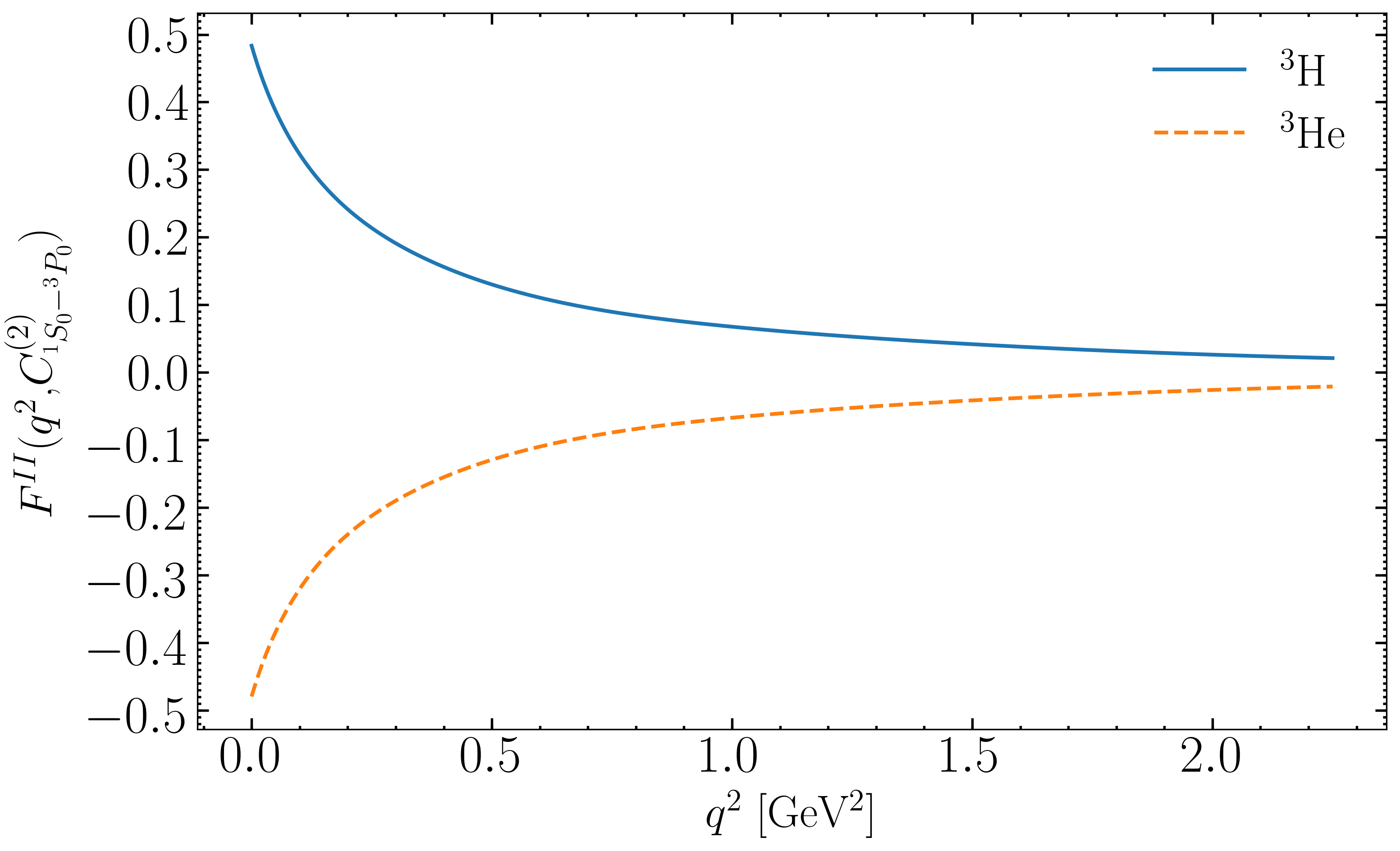}
}
  \caption{\label{fig:edffq2} The EDFF contributions arising from five
    different two-nucleon CP-odd operators as a function of $q^2$.     }
\end{figure}

\begin{figure}[t]
    \includegraphics[width = 0.9\textwidth]{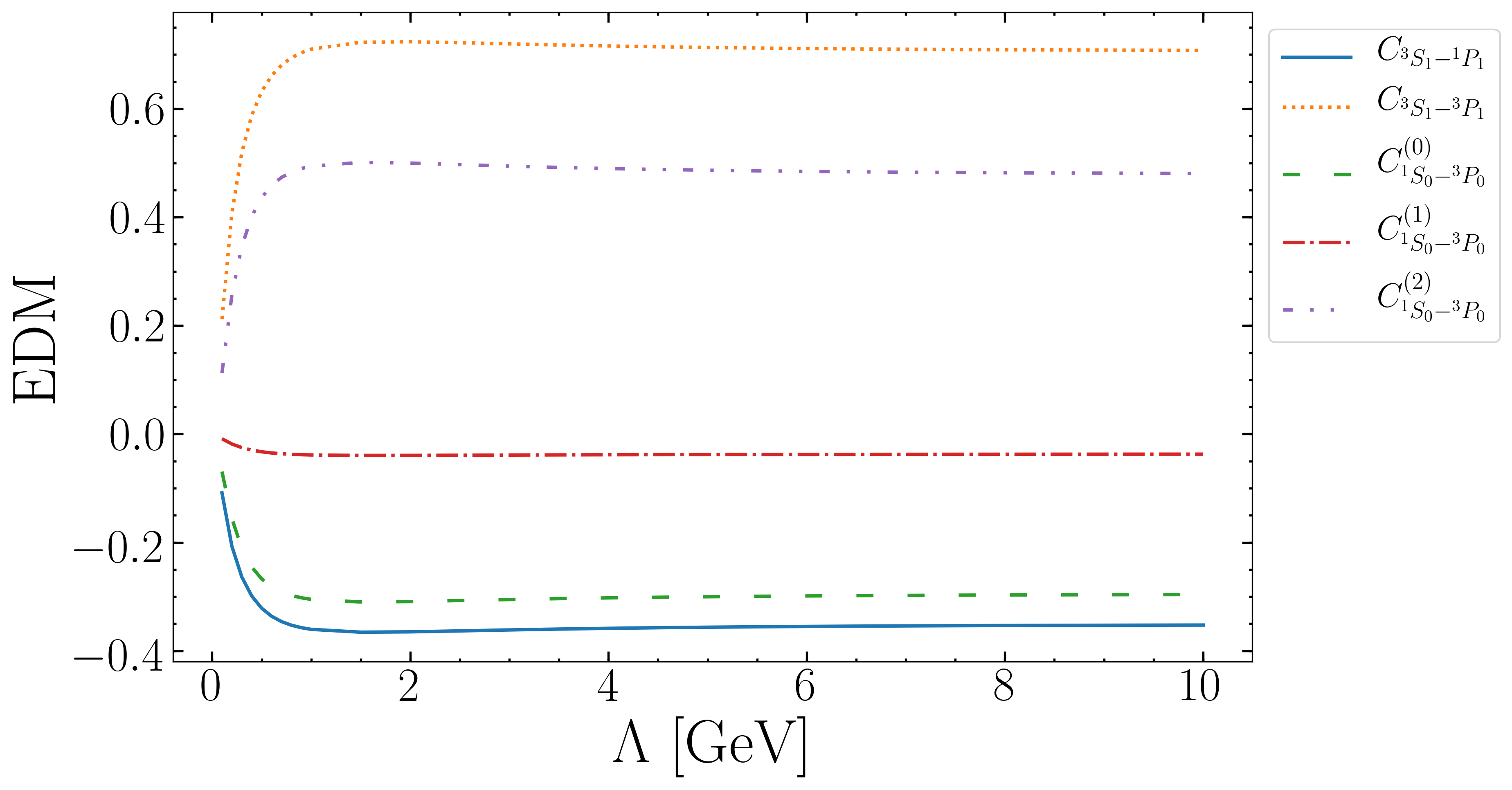}
  \caption{\label{fig:cutoff-dependence} Cutoff dependence of the ${}^3$H EDM two-nucleon contributions.}
\end{figure}

The EDFF results obtained for ${}^3$H are shown in
Fig.~\ref{fig:edffq2}. In Table~\ref{tbl:3H-2body-edm} we show the
cutoff dependence of the dipole moment contributions arising from the
different EFT operators. Furthermore,
we observe that cutoffs larger than $~1.5$~GeV are needed to obtain
numerically converged results. This convergence behavior is shown for
the EDMs in Fig.~\ref{fig:cutoff-dependence}.

At small $q^2$, we will expand the charge form factor as
\begin{eqnarray}
F_C(q^2) &=&   Z \left( 1  -  \frac{q^2}{6}  \langle r^2_{c} \rangle  +  \frac{1}{5!} \langle r^4_{c} \rangle q^4 + \ldots \right) ~,
\end{eqnarray}
where $r^2_c$ is the charge squared radius and $r^4_{c}$ the
4$^{\rm th}$ Zeemach moment, $Z$ denotes the total charge of the
nucleus considered and we omitted a label to denote the specific
nucleus.  We define a similar expansion for the one- and two-body
EDFF,
\begin{eqnarray}\label{twobodyFF}
F_{i}(q^2,C) &=&   d_i(C) \left( 1  -  \frac{q^2}{6}  \langle r^2_{d,\, i}(C) \rangle  +  \frac{1}{5!} \langle r^4_{d,\, i}(C) \rangle q^4 + \ldots \right),
\end{eqnarray}
where $i = \rm{I},\rm{II}$. $C= d_{n,p}$ for the one-body term, while it denotes one of the
nucleon-dimer T-odd operators in Eq. \eqref{eq:2nucleon-L-edm} for the
two-body contribution.  In the $SU(4)$ limit, all the dependence on
couplings factorizes into the universal function $\tilde F_{SU(4)}(q^2)$ and
a linear combination of low-energy constants, as shown  in
Eqs.~\eqref{eq:edff-2body} and \eqref{eq:edff-2bodyAPP}.
The square radius of the EDFF is particularly important since it determines the nuclear Schiff moment, and thus the EDMs of the atomic $^3$H and $^3$He species \cite{Schiff:1963zz}. 
More precisely, the Schiff moment is proportional to the difference of the charge and dipole radii \cite{deJesus:2005nb}
\begin{equation}\label{Schiff}
S_i(C) = - \frac{d_i(C)}{6} \left( \langle r^2_{d,\, i}( C) \rangle  - \langle r^2_c \rangle  \right)~,
\end{equation}
where again $i$ denotes either the one- (I) or two-body (II) contribution.

The three-nucleon charge form factor in EFT($\slashed{\pi}$) has
already been computed in
Refs. \cite{Platter:2005sj,Vanasse:2015fph,Vanasse:2017kgh}, including
next-to-leading order (NLO) and next-to-next-to-leading order
(N$^2$LO) corrections. At LO, and neglecting Coulomb interactions, one
finds
\begin{eqnarray}
  \langle r^2_{c}(^3{\rm H})  \rangle = 1.28 \, {\rm fm}^2~, \qquad 
  \langle r^2_{c}(^3{\rm He}) \rangle = 1.56\, {\rm fm}^2~.
\end{eqnarray}

These results are in
agreement with those in Refs.~\cite{Vanasse:2015fph,Vanasse:2017kgh}.
We will use the charge form factor as a point of comparison for the
momentum dependence of the EDFF.  In the $SU(4)$ limit,
\begin{align}
  \langle r^2_{c}(^3{\rm H})  \rangle &= \langle r^2_{c}(^3{\rm He})  \rangle = 1.32 \, {\rm fm}^2.
\end{align}
\begin{table}[t]
  \begin{tabular}{c||ccccccc}
    $\Lambda$ (GeV)  & $d_p$ & $d_n$ & $C_{^3S_1-^1P_1}$ & $C_{^3S_1-^3P_1}$ & $C_{^1S_0-^3P_0}^{(0)}$ & $C_{^1S_0-^3P_0}^{(1)}$ & $C_{^1S_0-^3P_0}^{(2)}$ \\
    \hline\hline
    10  & 0.982 & 0.008  & -0.358  & 0.708 & -0.297  & -0.038  & 0.481 \\
    30  & 0.988 & 0.010  & -0.356  & 0.706 & -0.295  & -0.037  & 0.479 \\
    80  & 0.990 & 0.010  & -0.358  & 0.708 & -0.297  & -0.038  & 0.481 \\
    600 & 0.991 & 0.010  & -0.359  & 0.708 & -0.298  & -0.038  & 0.481
  \end{tabular}
  \caption{\label{tbl:3H-2body-edm}Coefficients of the $^3$H EDM
    low-energy constants for different values of the cutoff $\Lambda$.}
\end{table}
The neutron and proton EDMs contributions to the $^3$H and $^3$He EDM are given by
\begin{eqnarray}\label{1body}
d_{\rm I}(^3{\rm H}) &=& 0.99 \, d_p +  9.7 \cdot 10^{-3} d_n, \qquad d_{\rm I}(^3{\rm He}) = 0.99 \, d_n +  9.7 \cdot 10^{-3} d_p.
\end{eqnarray}
The EDM only deviates by 1\% from the expectation in the Wigner
limit.  These results can be compared with chiral EFT calculations of
Ref. \cite{deVries:2011an,Bsaisou:2014zwa,Gnech:2019dod}. These
calculations include subleading effects in the strong potential, and
thus in the three-nucleon wavefunctions, and typically find the $d_p$
($d_n$) contribution to $^3$H ($^3$He) EDM to be roughly 10\% smaller
than Eq. \eqref{1body}.

The dominant momentum dependence of the EDFF is encoded by the dipole
square radius, which we find to be
\begin{eqnarray}
\langle r^2_{d,\, {\rm I}}(^3{\rm H}, d_n) \rangle  &=&   - 18.3 \, {\rm fm}^2,  \qquad 
                                                    \langle r^2_{d,\, {\rm I}}(^3{\rm H}, d_p) \rangle =    1.28 \, {\rm fm}^2, \\
\langle r^2_{d,\, {\rm I}}(^3{\rm He}, d_n) \rangle &=&  1.28 \, {\rm fm}^2, \qquad \langle r^2_{d,\, {\rm I}}(^3{\rm He}, d_p) \rangle =- 18.3 \,   {\rm fm}^2.  
\end{eqnarray}
The square radii agree very well with the triton charge radius.  This
has consequences for the Schiff moment, and thus the EDMs of atomic
$^{3}$He and $^3$H.
We see that in the case of $^3$H, the one-body Schiff moment vanishes
at LO in EFT($\slashed{\pi}$), $S_{\rm I}(^3{\rm H}, d_p) =0$. The one-body Schiff moment of $^3$He is small,
but non-vanishing,
\begin{equation}
S_{\rm I}(^3{\rm He}, d_n) =  \frac{d_n}{6} \left( 0.28 \right) \, {\rm fm}^2.
\end{equation}

We adopt the same expansion as Eq.~\eqref{twobodyFF} for the function
$F_{SU(4)}(q^2)$ and obtain for the two-body form factor in the
$SU(4)$ limit
\begin{equation}
d_{SU(4)} = -0.332, \qquad \langle r^2_d \rangle_{SU(4)} = 1.32 \, \textrm{fm}^2.
\end{equation}
where we used the average of spin-singlet and triplet binding momentum
and the triton binding energy in our calculation. In this limit, the momentum dependence
of the form factor seems to be dictated by the charge form factor, and
we find that, to very good approximation,
\begin{equation}
\frac{\tilde F_{SU(4)}(q^2)}{F_C(q^2)} = {\rm constant}.
\end{equation}
For $q$ between 0 and 500 MeV, this ratio deviates from a constant at
the per mille level. 
At the physical value of the scattering lengths,
the $^3$H and $^3$He EDMs from the two-body form factor $F_{\rm II}$
are given by
\begin{eqnarray}
d_{\rm II}(^3{\rm H})  &=& -0.358   C_{^3S_1 - ^1P_1}  + 0.707 C_{^3S_1 - ^3P_1} - 0.297 C^{(0)}_{^1S_0 - ^3 P_0} \nonumber \\
& & - 0.0368 C^{(1)}_{^1S_0 - ^3 P_0} + 0.480 C^{(2)}_{^1S_0 - ^3 P_0}, \\
d_{\rm II}(^3{\rm He})  &=&  0.358   C_{^3S_1 - ^1P_1}  + 0.707 C_{^3S_1 - ^3P_1} + 0.297 C^{(0)}_{^1S_0 - ^3 P_0} \nonumber \\
& & - 0.0375 C^{(1)}_{^1S_0 - ^3 P_0} - 0.480 C^{(2)}_{^1S_0 - ^3 P_0}.
\label{eq:d3He-2body}
\end{eqnarray}

where, as our central value, we took the EDFF at $\Lambda =
60$~GeV. As already remarked, the numerical accuracy is a the percent
level, and smaller than the LO EFT($\slashed{\pi}$) theoretical uncertainty.
The EDFF square radii induced by the CPV operators  in Eq. \eqref{eq:2nucleon-L-edm} 
are given in Table \ref{tbl:radii}.

We notice that, in the absence of the Coulomb interaction, the EDMs of
$^3$H and $^3$He follow simple isospin relations.  In particular, the
isoscalar and isotensor operators give rise to an isovector EDM, while
the isovector operators to an isoscalar three-nucleon EDM.  These
patterns can be understood by noticing that only the isovector piece
of the CP-even one-body electromagnetic current $J^0$ contributes to
$F_{\rm II}$ \cite{Liu:2004tq,Stetcu:2008vt,deVries:2011an}.

\begin{table}[t]
  \begin{tabular}{cc|| ccccc}
   &   & $C_{^3S_1-^1P_1}$ & $C_{^3S_1-^3P_1}$ & $C_{^1S_0-^3P_0}^{(0)}$ & $C_{^1S_0-^3P_0}^{(1)}$ & $C_{^1S_0-^3P_0}^{(2)}$ \\
    \hline\hline
   $ \langle r^2_{d,\, {\rm II}}(^3{\rm H}) \rangle$  & (fm$^2$) & 1.31 & 1.30          & 1.24 & 1.48  & 1.19\\
   $ \langle r^2_{d,\, {\rm II}}(^3{\rm He}) \rangle$ &(fm$^2$) & 1.90 & 1.50          & 1.83 & 4.58  & 1.19\\
    \end{tabular}
    \caption{\label{tbl:radii} Square radii of the two-body EDFF
      induced by the CPV operators in Eq. \eqref{eq:2nucleon-L-edm},
      computed at $\Lambda = 60$ GeV.  The $C_{^1S_0-^3P_0}^{(1)}$
      squared radii have a numerical error of approximately 10 \%
      since the corresponding dipole moments are relatively small. The
      other radii have few-percent numerical uncertainties, which we
      do not show. }
\end{table}

We observe that the EDM induced by the isoscalar operators
$C_{^3S_1 - ^1P_1}$ and $C^{(0)}_{^1S_0 - ^3P_0}$ and by the isospin-1
operator $C_{^3S_1 - ^3P_1}$ deviate from the $SU(4)$ limit by about
10\%. The EDM from the operators that connect the $^3S_1$ to $^1P_1$
and $^3P_1$ waves increases (in absolute value) by 10\%, while the EDM
from $C^{(0)}_{^1S_0 - ^3P_0}$ decreases by the same amount. We also
note that the isotensor operator $C^{(2)}_{^1S_0 - ^3P_0}$ shows a
larger, 30\% variation, from the Wigner limit.  Furthermore, we find
that, with the exception of $C^{(1)}_{^1S_0 - ^3P_0}$, all operators
induce matrix elements of $\mathcal O(1)$, as naively expected.  The
contribution of $C^{(1)}_{^1S_0 - ^3P_0}$ is suppressed by roughly a
factor of ten. It is interesting to note that in the case of $^3$H,
the momentum dependence of the form factor induced by
$C_{^3S_1 - ^1P_1}$, $C_{^3S_1 - ^3P_1}$ and $C^{(0)}_{^1S_0 - ^3P_0}$
cannot be distinguished from the charged form factor within
error. This leads to $S_{\rm II}$ being compatible with zero, implying that for
these operators the Schiff moment vanishes at LO in pionless EFT.
$C^{(2)}_{^1S_0 - ^3P_0}$ induces a non-zero, but small Schiff
moment. Specifically, in the case of $^3$He, all operators induce a
non-zero Schiff moment, but also in this case we expect subleading
corrections in EFT($\slashed{\pi}$) to be important.

\section{Summary}
\label{sec:summary}
In this work, we have shown that EFT($\slashed{\pi}$) is an efficient
framework that facilitates a straightforward calculation of the EDMs
and their corresponding form factors of three-nucleon systems. We
focused on the $^3$H and $^3$He systems at leading order in the
pionless EFT expansion and neglected Coulomb effects in the $^3$He
system. At this order, the only (CP-even) parameters that enter our
calculation are the deuteron binding energy, the two-nucleon
spin-singlet scattering length, and the three-body binding energy of
the state under consideration. Allowing for CP-odd interactions in the
few-nucleon sectors leads to a total of 7 parameters, where two of
them are the neutron and proton EDM and 5 arise from short-distance
physics in the two-nucleon system.

The deuteron and the isoscalar combination of the $^3$H and $^3$He
EDMs are mostly sensitive to the isovector coupling
$C_{^3S_1 - ^3P_1}$ (see App.~\ref{deuteron} for a derivation of the
EDFF and resulting EDM in pionless EFT). These two observables are thus
largely degenerate, and, neglecting the one-body piece, our
calculation finds
\begin{eqnarray}
\frac{d(^3{\rm H}) + d(^3{\rm He})}{2 d(^2 {\rm H})} =  0.71.
\end{eqnarray}
For comparison, the chiral EFT calculation of
Ref. \cite{Gnech:2019dod} finds the ratio to be between $0.77$ and
$0.80$, for both the isovector pion-nucleon coupling $\bar g_1$ and
for the linear combination $A_3 - A_4$, which corresponds to
$C_{^3S_1 - ^3 P_1}$.  The isovector combination
$d(^3{\rm H}) - d(^3{\rm He})$ probes the isoscalar couplings
$C_{^3S_1 - ^1P_1}$, $C^{(0)}_{^1S_0 - ^3P_0}$ and the isotensor
$C^{(2)}_{^1S_0 - ^3P_0}$, which are particularly important for the
QCD $\bar\theta$ term.  In chiral EFT, this linear combination cannot
be expressed only in terms of pion-nucleon CPV couplings, but requires
short-range nucleon-nucleon operators at LO \cite{deVries:2020loy}.

Specializing to the QCD $\bar\theta$-term, we combine
Eqs.~\eqref{1body} and \eqref{eq:d3He-2body} to obtain
\begin{eqnarray}\label{edmtheta}
  d_{^3{\rm He}}(\bar\theta) &=& d_n   + 0.358~C_{^3 S_1 - ^1P_1} +  0.297~C_{^1 S_0 - ^3P_0}~.
\end{eqnarray}
We can then use Eq.~\eqref{eq:theta_estimate} to write the above result
in terms of the dimensionless couplings $c_{^3 S_1 - ^1P_1}$ and
$ c^{(0)}_{^1S_0 - ^3 P_0} $ with size of order one 
\begin{eqnarray}
& &\simeq \left( 2.0    + 10\, c_{^3 S_1 - ^1P_1} +  8.6\, c^{(0)}_{^1S_0 - ^3 P_0}     \right) \cdot 10^{-3}  \, \bar\theta \, e \, \textrm{fm}~.
\end{eqnarray}
From Eq.~\eqref{edmtheta} we see that the $^3$He EDM can receive a
dominant two-body contribution, but of course more precise statements
require a first principle determination of the LECs.

Our approach does not facilitate an as direct identification of the
sources of possible non-zero EDMs in light nuclei as chiral EFT
does. However, it offers order-by-order renormalizability, a clear
understanding of the dependence of observables on the employed
ultraviolet regulator and exhibits the dependence of observables on
simple measurable two- and three-body observables such as the
effective range parameters. At, NLO the effective ranges in the
singlet and triplet channels and this correction will be of order
$\gamma_t\rho_t\approx 0.4$ where $\rho_t$ is the triplet effective
range. We note that Coulomb corrections can be included in
EFT$(\slashed{\pi})$ and are expected to give an approximately 10~\%
correction for $^3$He \cite{Konig:2015aka} and are thereby smaller
than the expected size of NLO range corrections.

Finally, we are optimistic that our EFT$(\slashed{\pi})$calculation can be
directly connected to QCD using lattice calculations, given recent
results obtained with lattice QCD for electroweak matrix
elements~\cite{Savage:2016kon,Davoudi:2020ngi} of two-nucleon system
and the possibility to carry out this calculation in a finite volume
\cite{Kreuzer:2010ti}.

\begin{acknowledgments}
  We acknowledge stimulating conversations with J.~de~Vries,
  U.~van~Kolck and R.~Talman.  This research has been funded by the
  National Science Foundation under Grant No.  PHY-1555030, by the
  U.S.  Department of Energy, Office of Science under Contract
  Nos. DE-AC05-00OR22725, DE-AC52-06NA25396 and DE-SC0019647, by the
  Department of Energy topical collaboration on ``Nuclear Theory for
  Double-Beta Decay and Fundamental Symmetries'' and by the Laboratory
  Directed Research and Development program of Los Alamos National
  Laboratory under project number 20190041DR.
\end{acknowledgments}

\begin{appendix}
 
\section{The deuteron  electric dipole form factor and EDM}
\label{deuteron}
\begin{figure}
\subfigure[]{
  \includegraphics[width = 0.4\textwidth]{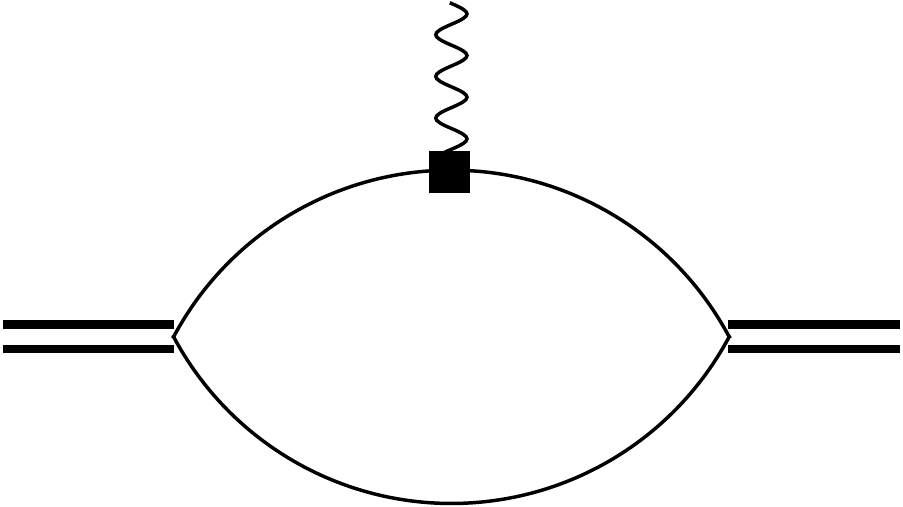}}
\subfigure[]{\includegraphics[width = 0.4\textwidth]{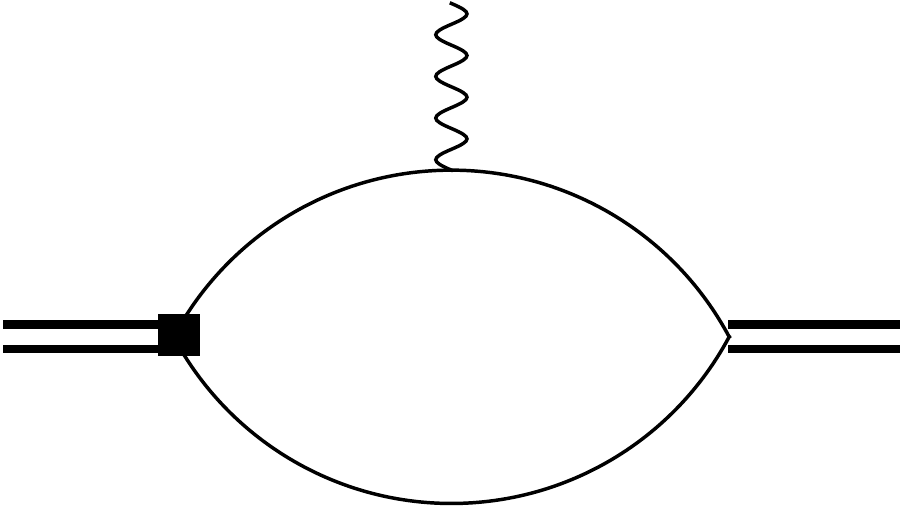}}
  \caption{Diagrams contributing to the deuteron EDM. The black
    squares denote insertions of CP-odd operators. We omitted the
    diagram that has the CP-odd operators to the right of the
    photon-nucleon vertex.}
  \label{fig:deuteronEDM}
\end{figure}

The diagrams that give the deuteron EDM are given in
Fig.~\ref{fig:deuteronEDM}.  The EDFF of the deuteron
$F(q^2,{}^2\hbox{H})$ is only sensitive to the isospin-one
$C_{^3S_1 - ^3P_1}$ coupling and we obtain
\begin{align}
  F(q^2,{}^2\hbox{H}) &=  \Big(d_n + d_p  - C_{^3S_1 -  ^3P_1} \Big)  \frac{4 \gamma_t}{q} \arctan \frac{q}{4\gamma_t}\\
  &= 
\left(d_n + d_p  - C_{^3S_1 -  ^3P_1} \right)F_c(q^2, {}^2\hbox{H})~,
\end{align}
here $F_c(q^2, {}^2\hbox{H})$ denotes the charge form factor of the
deuteron. The resulting EDM is obtained by taking the $q\rightarrow 0$
limit,
\begin{equation}
  d = \left(d_n + d_p  - C_{^3S_1 -  ^3P_1} \right)~.
\end{equation}
The direct proportionality of the EDFF to the charge form factor
causes the Schiff moment of the deuteron to be zero.

\section{Expressions for electric form factors diagrams $F_{C}$ and $F_{\rm I}$}
\label{App:Fcharge}
In this section, we give for completeness expressions for the diagrams
giving the CP-even electric form factor. To avoid confusion, we note
that we write $q = |\bm{q}|$. The three-nucleon wave function
renormalization $Z_{\psi}$ is defined as
\begin{equation}
  Z_{\psi} = \pi \left( \frac{ d \Sigma (E) }{ dE } \bigg|_{E = B} \right)^{-1}~,
\end{equation}
where the self-energy $\Sigma$ can be calculated via
\begin{equation}
  \Sigma (E) = -\pi  \widetilde{\mathbf{1}}^T \otimes_q \widetilde{ \bm{ \mathcal{G} } }(E,q)~.
\end{equation}

Below we give expressions for the contributions to the CP-even
three-nucleon form factor. The corresponding diagrams (a), (b), and
(c) are the same as Fig.~\ref{fig:one-body-diagrams}, but with a CP-even
photon vertex.

\paragraph*{\bf Diagram A:}
The calculation of the form factor diagrams is carried out in the
Breit frame. The vertex functions in diagram (a) that were originally
defined in the center-of-mass frame need to be boosted
as given in Eq. (\ref{eq:boostG}). This leads to the sum of the three
terms
\begin{equation}
  \label{eq:cp-even-diagram-a}
  F^{A}_{C \backslash \rm{I}}\left( q^2 \right)= 
  Z_{\psi}
    \Big[
      \mathcal{A}_{C \backslash \rm{I}}^{(a)} (q)
    +2 \widetilde{ \bm{ \mathcal{G} } }^{T}(E,p) \otimes_p \bm{ \mathcal{A} }^{(b)}_{C \backslash \rm{I}} (q, p)
    + \widetilde{ \bm{ \mathcal{G} } }^{T}(E,p) \otimes_p \bm{ \mathcal{A} }^{(c)}_{C \backslash \rm{I}} (q, p, k) \otimes_k \widetilde{
      \bm{ \mathcal{G} } }(E,k)
      \Big]~.
\end{equation}
For the CP-even and one-body CP-odd photon vertex, we define the following matrices,
\begin{align}
  \bm{M}_{C}^{A}
  = \begin{pmatrix}
    \frac{1+\tau_3}{2} & 0 \\
    0 &  \frac{3-\tau_3}{6}
  \end{pmatrix},~
  \bm{M}_{\rm I}^{A} 
  = 
  \frac{1}{6}
  \begin{pmatrix}
    d_n(\tau^3 - 1) - d_p(\tau^3 + 1) & 0 \\
     0 & d_n(\tau^3 + 3) - d_p(\tau^3 - 3)
   \end{pmatrix}.
\end{align}
The first term in Eq. (\ref{eq:cp-even-diagram-a}) is given by,
\begin{align}
  \nonumber
  \mathcal{A}_{C \backslash \rm{I}}^{(a)} (q)=&  \frac{M_N}{4 \pi^{2}} \Big|_0^1 \int_{0}^{\Lambda} d l  \int_{-1}^{1} d x \frac{l}{q x} 
  \widetilde{\mathbf{1}}^T
  \bm{M}_{C \backslash \rm{I}}^{A} 
  \bm{\mathcal{D}}(E, q, l, x, y)   \widetilde{\mathbf{1}}~.
\end{align}
where,
\begin{equation}
  \bm{ \mathcal{D} }(E, q, l, x, y) = 
  \bm{ D } 
  \left(E_{0}-\frac{l^{2}}{2 M_N}-\frac{q^{2}}{12 M_N}+ \big( \frac{1}{2}-y \big) \frac{l q x}{M_N}, l \right)~,
\end{equation}
and
\begin{equation}
  \Bigg|_{0}^{1} f(y)=f(1)-f(0)~.
\end{equation}

The second term of Eq. (\ref{eq:cp-even-diagram-a}) includes the CP-even vertex function and the function $\mathcal{A}_{C \backslash \rm{I}}^{(b)}(q, p)$ that is defined as

\begin{multline}
  \bm{ \mathcal{A} }_{C \backslash \rm{I}}^{(b)} (q, p)= 
  \frac{M_N}{2 \pi} \Big|_0^1 \int_{0}^{\Lambda} d l \int_{-1}^{1} d x \frac{l}{q x} \frac{1}{ p \sqrt{ l^2- \frac{2}{3} l q x + \frac{1}{9} q^2 } }  
  Q_{0}^{B}(q,l,p,x,y,2) \\
  \times
  \left(\begin{matrix}
    -1 & \phantom{+}3 \\
    \phantom{+}3 & -1
  \end{matrix}\right)
  \bm{M}_{C \backslash \rm{I}}^{A} 
    \bm{ \mathcal{D} }(E, q, l, x, y)
    \widetilde{\mathbf{1}}~,
\end{multline}

where, $x$ is defined through $\bm{l} \cdot \bm{q} = lqx$. We defined also a boosted version of the function $Q_0$
\begin{equation} \label{Q0_boost}
  Q_{0}^{boost}(q,l,k,x,y,z) = Q_0\left( \frac{k^2 + l^2 + \frac{q^2}{9} + (y - \frac{z}{3}) lqx -M_N B}{ k \sqrt{l^2 + \frac{q^2}{9} - (-1)^z \frac{2}{3} lqx }}\right)~.
\end{equation}

And the third term in Eq. (\ref{eq:cp-even-diagram-a}) includes the following function

\begin{align}
  \nonumber
  \bm{ \mathcal{A} }^{(c)}_{C \backslash \rm{I}}  (q, p, k) = & 
  M_N \Big|_0^1 \int_{0}^{\Lambda} d l \int_{-1}^{1} d x \frac{l}{q x} 
  \frac{  Q_{0}^{B}(q,l,k,x,y,1) Q_{0}^{B}(q,l,p,x,y,2) }
  {k p \sqrt{l^{2}+\frac{2}{3} lq x+\frac{1}{9} q^{2}} 
  \sqrt{l^{2}-\frac{2}{3} lq x+\frac{1}{9} q^{2}}} 
  \\
  & 
  \times \left(\begin{matrix}
    -1 & \phantom{+}3 \\
    \phantom{+}3 & -1
  \end{matrix}\right)
  \bm{M}_{C \backslash \rm{I}}^{A} 
  \bm{ \mathcal{D} }(E, q, l, x, y)
  \left(\begin{matrix}
    -1 & \phantom{+}3 \\
    \phantom{+}3 & -1
  \end{matrix}\right) ~.
\end{align}

\paragraph*{\bf Diagram B:}
For the CP-even and one-body CP-odd photon vertex, we define the following matrices,
\begin{align}
  \bm{M}_{C}^{B}
  =&
  \begin{pmatrix}
    \frac{\tau^3 - 1}{2} 
    & \frac{3 + \tau^3}{2} \\
    \frac{3 + \tau^3}{2}
    & - \frac{3 + 5\tau^3}{6}
\end{pmatrix},~
\nonumber
\\
  \bm{M}_{\rm I}^{B} 
  =&
  \frac{1}{12}
  \begin{pmatrix}
   5d_p(\tau^3 - 1) - 5 d_n(\tau^3 + 1)
        &   d_p(\tau^3 + 3) - d_n(\tau^3 - 3) \\
        d_p(\tau^3 + 3) - d_n(\tau^3 - 3) 
        & d_p(5\tau^3 + 3) - d_n(5\tau^3 -3) 
    \end{pmatrix}.
\end{align}

The contribution from diagram (b) in Fig.~\ref{fig:one-body-diagrams}  is given by
\begin{equation}
  F^{B}_{C \backslash \rm{I}}\left( q^2 \right)= 
  Z_{\psi} \int_{-1}^{1} dx \int_{-1}^{1} dy \, 
    \widetilde{ \bm{ \mathcal{G} } }^{T}(E,p) \otimes_p \Gamma^{B} (q, p, k)
    \bm{M}_{C \backslash \rm{I}}^{B} 
   \otimes_k \widetilde{ \bm{ \mathcal{G} } }(E,k)~,
\end{equation}
where we defined
\begin{align} 
  \nonumber
  \label{eq:gammab}
  \Gamma^{B}(q,p,k,x, y) &=
  -\frac{M_N}{4}
 \int_0^{2\pi} d\phi  
  \\
  \nonumber
  &
 \times \left[ k^2 + p^2 
  + kp( xy + \sqrt{1-x^2}\sqrt{1-y^2} \cos \phi)
  +\frac{1}{3}q(kx + 2py)
  + \frac{1}{9} q^2 -M_N B \right]^{-1}
  \\ &
 \times \left[ k^2 + p^2 
  + kp( xy + \sqrt{1-x^2}\sqrt{1-y^2} \cos \phi)
  -\frac{1}{3}q(2kx + py)
  + \frac{1}{9} q^2 -M_N B \right]^{-1}~,
\end{align}
where the x-, y- and $\phi$-integrals are angular integrals
\begin{align}
  \mathbf{k} \cdot \mathbf{q} &= kqx~,  \label{eq: diagramB_x}
  \\
  \mathbf{p} \cdot \mathbf{q} &= pqy~, \label{eq: diagramB_y}
  \\
  \mathbf{k} \cdot \mathbf{p} &= k p \cos \phi .
\end{align}

\paragraph*{\bf Diagram C:}
Finally, for the CP-even and one-body CP-odd photon vertex, we define the following matrices,
\begin{align}
  \bm{M}_{C}^{C}
  =
  \begin{pmatrix}
    1 &  0\\
    0 & 1 + \frac{2 \tau^3}{3}
  \end{pmatrix},~
  \bm{M}_{\rm I}^{C} 
  =
  \begin{pmatrix}
    2\tau^3(d_n + d_p) & d_p - d_n  \\
    d_p - d_n & 0
  \end{pmatrix}
.
\end{align}

The contribution from diagram (c) in Fig.~\ref{fig:one-body-diagrams} is given by
\begin{equation}
  F^{C}_{C \backslash \rm{I}}\left( q^2 \right)= 
  Z_{\psi}
\int_{-1}^{1}dx\,  \Gamma^C(q, k,x) \otimes_k 
  \left[ \widetilde{  \bm{ \mathcal{G} } }^{T}(E,p) 
  \bm{M}_{C \backslash \rm{I}}^{C}
   \widetilde{  \bm{ \mathcal{G} } }(E,k) \right]~,
\end{equation}
where $x$ is defined through $ {\bf k} \cdot {\bf q} = kqx $ and the function $\Gamma^C(q,k)$ is defined as

\begin{align} 
  \label{eq:gammac}
  \Gamma^C(q,k,x) = 
  \frac{M_N}{q} 
  \arctan\left( \frac{q}{ 2\sqrt{\frac{3}{4} p^2 -M_N B} + 2\sqrt{\frac{3}{4} k^2 -M_N B} }\right)~,
\end{align}
and,
\begin{align}
    {\bf p} =& {\bf k} + \frac{1}{3}{\bf q}~. \label{eq: diagramC_p}
\end{align}

\section{Expressions for form factor diagrams $F_{\rm II}$}
\label{App:FII}
Below we give the expression for the contributions to the CP-odd form
factor arising from CP-odd two-nucleon operators.

\subsection{Boosted Vertex functions ( of Diagram A)}
The calculation of the form factor diagrams is carried out in the
Breit frame. This requires us to relate the vertex functions that were
defined in the center of mass frame to boosted ones. To do this we use
the following integrals that gives us the boosted CP-even vertex
function $\mathcal{G}$ (see also Ref.~\cite{Vanasse:2015fph}):
\begin{equation}
  \label{eq:boostG}
  \bm{ \mathcal{G} }^{\rm boost} (q,l,x,y,z) = 
  \widetilde{\mathbf{1}}
  + 
  R_{0}^{\rm boost} (q,l,k,x,y,z)
\begin{pmatrix}
  - 1 &  \phantom{+}3 
    \\
    \phantom{+}3 & -1
  \end{pmatrix} 
  \otimes_k
  \bm{ \tilde{\mathcal{G}} }(E,k)~,
\end{equation}
here $x$ denotes the cosine of the angle between the boost momentum
$q$ and the relative momentum $l$ between the dimer and the nucleon
field. We have also already carried out the $l_0$ loop integration
that enters when the vertex functions is folded with remaining
parts of the diagrams for the matrix elements. The factor $z$ is
introduced for convenience to have a short-hand notation for the
kinematically different vertex functions on the left or right hand
side of the photon vertex.

We also need the boosted CP-odd vertex function $\mathcal{T}$:
\begin{multline}
  \label{eq:CP-oddVF}
  \bm{ \mathcal{T} }_{A}^{\rm boost} (q,l,x,y,z) = 
  \left(\frac{lx}{q} +\- \frac{(-1)^{ z  } }{3} \right)
  \left[
  \left( \begin{matrix}
    \frac{1 + \tau^3 }{2} & 0
    \\
    0 & \frac{3 - \tau^3}{6}
  \end{matrix}\right)
  \bm{ \mathcal{T} }^{\frac{1}{2}, {\rm boost} } (q,l,x,y,z)\right.\\
  +\left.
  \left( \begin{matrix}
    0 & 0
    \\
    0 & \frac{2}{3}
  \end{matrix}\right)
  \bm{ \mathcal{T} }^{\frac{3}{2},{\rm boost}} (q,l,x,y,z)
  \right]~.
\end{multline}
Here, for simplicity, we defined two boosted functions
$ R_{0}^{\rm boost} $ and $ R_{1}^{\rm boost}$ with
$Q_{0}^{\rm boost}$ defined in Eq.(\ref{Q0_boost})
\begin{align}
  R_{0}^{\rm boost}(q,l,k,x,y,z) =& \frac{2\pi}{ k \sqrt{l^2 + \frac{q^2}{9} - (-1)^z \frac{2}{3} lqx} } Q_{0}^{\rm boost}(q,l,k,x,y,z)~,
  \\
  R_{1}^{\rm boost}(q,l,k,x,y,z) =& \frac{2\pi}{  l^2 + \frac{q^2}{9} - (-1)^z \frac{2}{3} lqx }   
  \nonumber \\
   & \times \left( 1 -\frac{k^2 + l^2 + \frac{q^2}{9} + (y - \frac{z}{3}) lqx - M_N B}{ k \sqrt{l^2 + \frac{q2}{9} - (-1)^z\frac{2}{3} lqx }}  Q_{0}^{\rm boost}(q,l,k,x,y,z) \right)~.
\end{align}
The boosted isospin-projected CP-odd vertex functions required in Eq.(\ref{eq:CP-oddVF}) are given by
\begin{align}
  \nonumber
   \bm{ \mathcal{T} }^{ \frac{1}{2}, {\rm boost}} (q,l,x,y,z)  =&
  \bm{R}^{ \frac{1}{2},{\rm boost}}_{\mathcal{T}} (q,l,k,x,y,z)
 \otimes_k  
  \widetilde{ \bm{ \mathcal{T} } }^{\frac{1}{2}} (E,k) 
  + \bm{R}^{ \frac{1}{2},{\rm boost}} (q,l,k,x,y,z)
  \otimes_k  \bm{ \widetilde{\mathcal{G}} }(E,k) ~,
  \\
  \bm{ \mathcal{T} }^{ \frac{3}{2},{\rm boost}}(q,l,x,y,z) =&
  \bm{R}^{ \frac{3}{2},{\rm boost}}_{\mathcal{T}} (q,l,k,x,y,z)
  \otimes_k  
\widetilde{ \bm{ \mathcal{T} } }^{\frac{3}{2}} (E,k) 
+ \bm{R}^{ \frac{3}{2},{\rm boost} } (q,l,k,x,y,z)
\otimes_k  \bm{ \widetilde{\mathcal{G}} }(E,k) ~,
\end{align}
where, 
\begin{align}
  \nonumber
  \bm{ R }^{ \frac{1}{2},{\rm boost}}_{\mathcal{T}} (q,l,k,x,y,z)
  =&     \tilde{R}_{1}^{\rm boost}(q,l,k,x,y,z)
  \left( \begin{matrix}
    -1 & \phantom{+}3 \\
    \phantom{+}3 & -1
  \end{matrix} \right),
  \\
  \nonumber
  \bm{ R }^{ \frac{3}{2} ,{\rm boost}}_{\mathcal{T}} (q,l,k,x,y,z)
  =&      \tilde{R}_{1}^{\rm boost}(q,l,k,x,y,z)
  \left( \begin{matrix}
    0 & 0 \\
    0 & 2
  \end{matrix} \right),
  \\
  \nonumber
  \bm{ R }^{ \frac{1}{2},{\rm boost} } (q,l,k,x,y,z)
  = &    
 \tilde{R}_{0}^{\rm boost}(q,l,k,x,y,z)   \Bigg[
  \left( \begin{matrix}
    -1 & 1 \\
    -2 & 0
  \end{matrix} \right)
   \left( C_{^3S_1 - ^1P_1} + \frac{2}{3} \tau^3 C_{^3S_1 - ^3 P_1} \right)   \nonumber  \\
&    + \left( \begin{matrix}
  \phantom{+}0 & 2 \\
    -1 & 1
  \end{matrix} \right)  
  \left( C^{(0)}_{^1S_0 - ^3P_0} - \frac{2}{3} \tau^3 C^{(1)}_{^1S_0 - ^3 P_0} \right) \Bigg]
  \nonumber
  \\
 &  +  \tilde{R}_{1}^{\rm boost}(q,l,k,x,y,z) \Bigg[
  \left( \begin{matrix}
    \phantom{+}1 & 2 \\
    -1 & 0
  \end{matrix} \right)
   \left( C_{^3S_1 - ^1P_1} + \frac{2}{3} \tau^3 C_{^3S_1 - ^3 P_1} \right) 
     \nonumber  \\
&    + \left( \begin{matrix}
  \phantom{+}0 & \phantom{+}1 \\
    -2 & -1
  \end{matrix} \right)  
  \left( C^{(0)}_{^1S_0 - ^3P_0} - \frac{2}{3} \tau^3 C^{(1)}_{^1S_0 - ^3 P_0} \right) \Bigg]~,
\nonumber
  \\
  \nonumber
  \Big( \bm{ R }^{ \frac{3}{2},{\rm boost} } \Big)^T (q,l,k,x,y,z)
  = &     \tilde{R}_{0}^{\rm boost}(q,l,k,x,y,z)
  \frac{1}{3}\left( \begin{matrix}
    0 &  8 C^{}_{^3S_1 - ^3P_1} - C^{(1)}_{^1S_0 - ^3 P_0} + 3 \tau_3 C^{(2)}_{^1S_0 - ^3 P_0}
    \\
    0
    & 
    -5 (C^{(1)}_{^1S_0 - ^3 P_0} - 3 \tau_3 C^{(2)}_{^1S_0 - ^3 P_0})
  \end{matrix} \right) 
  \\
  & +
  \tilde{R}_{1}^{\rm boost}(q,l,k,x,y,z) 
  \frac{1}{3}\left( \begin{matrix}
    0 &  4C_{^3S_1 - ^3P_1} - 2 \left(C^{(1)}_{^1S_0 - ^3 P_0} - 3 \tau_3 C^{(2)}_{^1S_0 - ^3 P_0} \right)
    \\
    0
    & 
    -4 \left(C^{(1)}_{^1S_0 - ^3 P_0} - 3 \tau_3 C^{(2)}_{^1S_0 - ^3 P_0} \right)
  \end{matrix} \right) ~.
\end{align}
\subsection{Diagram A}
Diagram (a) in Fig.~\ref{fig:two-body-diagrams} is given by
\begin{multline}
  F^{A}_{\rm II}(q^2) =    Z_{\psi}
  \Bigg|_0^1 \int_0^\Lambda dl \int_{-1}^1 dx 
  \frac{M_N}{4\pi^2} \frac{2}{3} \frac{l}{qx}
  \Bigg[ 
  \left( \bm{ \mathcal{G} }^{\rm boost} \right)^T (q,l,x,y,2)
  \bm{ \mathcal{D} } (E, q, l, x, y)
  \bm{  \mathcal{T} }_A^{\rm boost} (q,l,x,y,1)
  \\
   -
  \left( \bm{ \mathcal{T} }_A^{\rm boost} \right)^T (q,l,x,y,2)
  \bm{ \mathcal{D} }(E, q, l, x, y)
  \bm{ \mathcal{G} }^{\rm boost} (q,l,x,y,1) \Bigg]~,
\end{multline}
where the boosted CP-even and CP-odd vertex functions were defined above.

\subsection{Diagrams B and D}
\paragraph*{\bf Diagram B:}
Diagram (b) in Fig.~\ref{fig:two-body-diagrams} is given by
\begin{align}
  \nonumber
  F^B_{\rm II}(q^2) = &  Z_{\psi}
  \frac{1}{q^2} \int_{-1}^{1} dx \int_{-1}^{1} dy \, \Bigg\lbrace
  \widetilde{ \bm{ \mathcal{G} } }(E,p) \otimes_p \bigg( \Gamma^{B}(q,p,k, x, y)\, \mathbf{k} \cdot \mathbf{q}  \bigg) \otimes_k 
  \widetilde{ \bm{ \mathcal{T} }  }_{B} (E, k )
  \\
  &
  -
  \widetilde{ \bm{ \mathcal{T} } }_{B}(E,p) \otimes_p \bigg( 
    \mathbf{p} \cdot \mathbf{q} \,\Gamma^{B} (q,p,k,x,y) \bigg) \otimes_k 
  \widetilde{ \bm{ \mathcal{G} } } (E, k ) \Bigg\rbrace~,
\end{align}
where,
\begin{align}
  \widetilde{ \bm{ \mathcal{T} }  }_B (E, k )
  = & 
  \left( \begin{matrix}
    \frac{\tau^3 - 1 }{2} 
    &
    \frac{\tau^3 + 3 }{2} 
    \\
    \frac{\tau^3 + 3 }{2} 
    &
    \frac{-5\tau^3 - 3 }{6} 
  \end{matrix}\right)
  \bm{ \widetilde{ \mathcal{T} } }^{ \frac{1}{2} } (E, k) +
  \left( \begin{matrix}
    0 &-2 \\
    0 & -\frac{2}{3}
  \end{matrix}\right)
  \bm{ \widetilde{ \mathcal{T} } }^{ \frac{3}{2} }(E, k)~.
\end{align}

\paragraph*{\bf Diagram D:}
Diagram (d) in Fig.~\ref{fig:two-body-diagrams} can be written as
\begin{multline}
  F^D_{\rm II}(q^2) = 
  Z_{\psi} \int_{-1}^{1} dx \int_{-1}^{1} dy \, 
  \widetilde{ \bm{ \mathcal{G} } }(E,p) \otimes_p \Bigg\lbrace \Gamma^{B} (q,p,k,x,y)\\
\times  \left[ \frac{ kx +  2py - \frac{2}{3}q }{q} \bm{M}_D^T 
  - \frac{ 2kx +  py + \frac{2}{3}q }{q} \bm{M}_D \right]
  \Bigg\rbrace
  \otimes_k \widetilde{ \bm{ \mathcal{G} } }(E,k)~,
\end{multline}
where,
\begin{align}
  \bm{M}_D = 
  \left( \begin{matrix}
    \frac{\tau^3 - 1}{6} ( 3 C_{^3S_1-^1P_1} - 2C_{^3S_1-^3P_1} )
    &\quad
     \frac{3 + \tau^3 }{6}C^{(0)}_{^1S_0-^3P_0} 
    - \frac{1 + \tau^3 }{3} C^{(1)}_{^1S_0-^3P_0}
    + \frac{2\tau^3}{3} C^{(2)}_{^1S_0-^3P_0}
    \\
    -\frac{3+\tau^3}{6}\left( C_{^3S_1-^1P_1}  + 2C_{^3S_1-^3P_1}\right)
    &\quad
    \frac{3 + 5\tau^3 }{6} C^{(0)}_{^1S_0-^3P_0} 
    - \frac{1 + \tau^3 }{3} C^{(1)}_{^1S_0-^3P_0}
    - \frac{2\tau^3}{3}C^{(2)}_{^1S_0-^3P_0}
  \end{matrix}\right)~.
\end{align}

Again, recall that $x$, $y$ and the function $\Gamma^B (q,p,k,x,y) $ are defined in Eq.~\eqref{eq: diagramB_x}, Eq.~\eqref{eq: diagramB_y} and Eq.~\eqref{eq:gammab}.
\subsection{Diagrams C and E}
\paragraph*{\bf Diagram C:}
Diagram (c) in Fig.~\ref{fig:two-body-diagrams} leads to
\begin{align}
  F^C_{\rm II}(q^2) =&
  Z_{\psi}\int_{-1}^{1}dx\,
  \Gamma^{C}(q,k,x) \otimes_k \frac{1}{q^2}
  \Big[
  ({\bf k} \cdot {\bf q})  \widetilde{ \bm{ \mathcal{G} } }^{T} (E,p) \widetilde{ \bm{ \mathcal{T} } }_C(E,k)
  -( {\bf p} \cdot {\bf q})
                       \widetilde{ \bm{ \mathcal{T} } }_C^{T} (E,p)  \widetilde{ \bm{ \mathcal{G} } }(E,k)
  \Big]~,
\end{align}
where,
\begin{align}
     \widetilde{ \bm{ \mathcal{T} } }_C (E,k) & = 
     \left( \begin{matrix}
       1 & 0 \\
       0 & 1 + \frac{2}{3} \tau^3
     \end{matrix}\right)
     \bm{ \widetilde{ \mathcal{T} } }^{ \frac{1}{2} } (E,k)
      +
      \left( \begin{matrix}
        0 & \phantom{+}0 \\
        0 & -\frac{2}{3} 
      \end{matrix}\right)
      \bm{ \widetilde{ \mathcal{T} } }^{ \frac{3}{2} } (E,k)~.
\end{align}

\paragraph*{\bf Diagram E:}

Diagram (e) in Fig.~\ref{fig:two-body-diagrams} is given by
\begin{align}
  \nonumber
  F^E_{\rm II} (q^2) =&  2 Z_{\psi} \int_{-1}^{1}dx\,
   \Gamma^{C}(q,k,x)  \otimes_k 
  \widetilde{ \bm{ \mathcal{G} } }^{T} (E,p)
  \Bigg[ \frac{{\bf p} \cdot {\bf q} + \frac{q^2}{3} }{q^2} 
  \\
  & + \frac{1}{2q} \left( \sqrt{- M_N B + \frac{3}{4}k^2} - \sqrt{- M_N B + \frac{3}{4}p^2}\right)  \Bigg]  \bm{M}_E
  \widetilde{ \bm{ \mathcal{G} } }(E,k) ~,
\end{align}

and we define the matrix $\bm{M}_E$
\begin{equation}
  \bm{M}_E
  =
  \frac{\tau^3 }{3} 
  \left( \begin{matrix}
  2\tau^3 C_{^3S_1-^3P_1}
    &
 C_{^3S_1-^1P_1}
    \\
    - ( C^{(0)}_{^1S_0-^3P_0} -2C^{(2)}_{^1S_0-^3P_0} )
    &
    0
  \end{matrix}\right)~.
\end{equation}

Recall that $x$, $p$ and the function $\Gamma^C(q, k, x) $ are defined
previously above and in Eq.(\ref{eq: diagramC_p}), and
Eq. (\ref{eq:gammac}).

\section{Expressions for form factors diagrams $F_{SU(4)}$}\label{App:FII_SU4}

In the SU(4) limit, the two-body diagrams could be simplified to a universal
function depends on $q$ times a combination of T-odd coefficients.
\begin{eqnarray}
  \label{eq:edff-2bodyAPP}
F_{\rm II}(q^2, SU(4))  =  \widetilde{F}_{SU(4)}(q^2) 
    \left(
    \tau^3  C^{}_{^3S_1 - ^1P_1}
    + 2  C^{}_{^3S_1 - ^3P_1}
    + \tau^3  C^{(0)}_{^1S_0 - ^3P_0}  
    - 2 \tau^3  C^{(2)}_{^1S_0 - ^3P_0}  
    \right), 
\end{eqnarray}

Thus, the universal electric dipole form factors also has five terms,
\begin{align}
  \widetilde{F}_{SU(4)}(q^2) =&
  \widetilde{F}_{SU(4),A}( q^2 )+  \widetilde{F}_{SU(4),B}( q^2 )+  \widetilde{F}_{SU(4),C}( q^2 )+  \widetilde{F}_{SU(4),D}( q^2 )
  +  \widetilde{F}_{SU(4),E}( q^2 )
\end{align}

The first term is given by,
\begin{align}
  \widetilde{F}_{SU(4),A}( q^2 ) = &   Z_{\psi}
  \Bigg|_0^1 \int_0^\Lambda dl \int_{-1}^1 dx 
  \frac{ M_N }{4\pi^2} \frac{4l   \mathcal{D}_+  (E, q, l, x, y) }{9qx}
  \Bigg[ 
  \mathcal{G}_+^B  (q,l,x,y,2)
  \mathcal{T}_{SU(4)} (q,l,x,y,1)
  \left(\frac{lx}{q} - \frac{1}{3} \right)
  \nonumber \\
  & -
  \left(\frac{lx}{q} + \frac{1}{3} \right)
  \mathcal{T}_{SU(4)} (q,l,x,y,2)
\mathcal{G}_+^B (q,l,x,y,1) \Bigg]~,
\end{align}
where the boosting is carried out analogously to Appendix \ref{App:FII}.

The second term is given by,
\begin{align}
  \widetilde{F}_{SU(4),B}(q^2) = &  Z_{\psi}
  \frac{1}{q^2} \int_{-1}^{1} dx \int_{-1}^{1} dy \,
  \frac{4}{3} \Bigg\lbrace
  \widetilde{ \mathcal{G} }_+(E,p) \otimes_p \bigg( 
    \Gamma^{B}(q,p,k,x,y) \mathbf{k} \cdot \mathbf{q}   \bigg) \otimes_k 
  \widetilde{  \mathcal{T}  }_{ SU(4) } (E, k )
  \nonumber \\
  &
  -
  \widetilde{  \mathcal{T}  }_{SU(4)} (E,p) \otimes_p \bigg( 
    \mathbf{p} \cdot \mathbf{q} \Gamma^{B} (q,p,k,x,y) \bigg) \otimes_k 
  \widetilde{ \mathcal{G}}_+ (E, k ) \Bigg\rbrace ~,
\end{align}

where $\mathcal{G}_+ (E, k)$ and $ \mathcal{T}_{SU(4)} $ are defined in
Eq. (\ref{su4_g}) and Eq. (\ref{su4_7}).

Similarly, we give the remaining terms,
\begin{multline}
  \widetilde{F}_{SU(4),C}(q^2) =
  Z_{\psi}\int_{-1}^{1} dx\,
  \Gamma^{C}(q,k,x) \otimes_k \frac{1}{q^2}  \frac{-2}{3}
  \Big[
  ({\bf k} \cdot {\bf q})  \widetilde{  \mathcal{G}  }_+ (E,p) \widetilde{ \mathcal{T} }_{SU(4)} (E,k)\\
  -( {\bf p} \cdot {\bf q})
  \widetilde{  \mathcal{G}  }_+ (E,p)  \widetilde{ \mathcal{T} }_{SU(4)}  (E,k)
  \Big]~,
\end{multline}

\begin{align}
  \nonumber
  \widetilde{F}_{SU(4),D}(q^2) =&
  -\int_{-1}^{1} dx \int_{-1}^{1} dy \,
  \widetilde{\mathcal{G}}_+(E,p) \otimes_p     
  \Gamma^{B}(q,p,k,x,y) \otimes_k \widetilde{\mathcal{G}}_+(E,k)\\
&\times \frac{2}{3} \frac{ kx -  py + \frac{4}{3}q }{q} ~,
\end{align}

and
\begin{align}
  \widetilde{F}_{SU(4),E}(q^2) =&   Z_{\psi}\int_{-1}^{1} dx
  \frac{1}{3} \Gamma^{C}(q,k,x)  \otimes_k 
  \widetilde{ \mathcal{G}}_+ (E,p)
  \widetilde{ \mathcal{G}}_+(E,k)~.
\end{align}

Recall that $\Gamma^{B}(q,p,k,x,y)$, $\Gamma^{C}(q,k,x)$ and other variables are
defined previously in the corresponding subsections in Appendix \ref{App:FII}.

\end{appendix}

\end{document}